\documentclass[twocolumn]{aastex631}

\begin{document}

\newcommand{\degg}{$^{\circ}$}

\title{Extinction values towards embedded planets in protoplanetary disks \\ estimated from hydrodynamic simulations}

\correspondingauthor{Felipe Alarc\'on}
\email{falarcon@umich.edu}

\author[0000-0002-2692-7862]{Felipe Alarc\'on }
\affiliation{Department of Astronomy, University of Michigan,
323 West Hall, 1085 S University, Ave.,
Ann Arbor, MI 48109, USA}

\author[0000-0003-4179-6394]{Edwin A. Bergin}
\affiliation{Department of Astronomy, University of Michigan,
323 West Hall, 1085 S University, Ave.,
Ann Arbor, MI 48109, USA}

\author[0000-0001-7255-3251]{Gabriele Cugno}
\affiliation{Department of Astronomy, University of Michigan,
323 West Hall, 1085 S University, Ave.,
Ann Arbor, MI 48109, USA}



\begin{abstract}

The upcoming new coronographs with deeper contrast limits, together with planned and current high-contrast imaging campaigns will push the detectability limit of protoplanets. These planet-hunting campaigns present a new opportunity to characterize protoplanets and their surrounding environments.  However, there are clear uncertainties as to what are the extinction levels at different regions of protoplanetary disks, which will impede our ability to characterize young planets. A correct understanding of the expected extinction together with multiple photometric observations will lead to constraints on the extinction levels, dust growth, disk evolution and protoplanetary accretion rates. In this work, we used hydrodynamic simulations and protoplanetary disk observational constraints obtained from both dust and gas emission to explore the expected extinction maps for continuum filters associated with strong hydrogen lines as tracers of accretion and key broadband photometric filters. We provide a scaling relationship for the extinction as a function of planetary separation and disk mass for three different gas giant masses. We also report values for a subset of disks of interest targetted by multiple imaging campaigns.  The described values will be useful for the optimal design of future planet-hunting surveys and for giving context to non-detections in protoplanetary disks and the observed fluxes of point sources along with the birth conditions of protoplanets.
\end{abstract}


\section{Introduction} \label{sec:intro}

As dedicated algorithms for high-contrast imaging are being applied in planet-hunting campaigns, the detection of protoplanets through direct methods has started to become a reality \citep[e.g.][]{Keppler..2018, Haffert..et..al..2019, Currie..2022,Yifan..2022, Hammond..2023}. However, many dedicated surveys have shown that the detection rate of planetary companions in circumstellar disks is quite low \citep[][among others]{Cugno2019, Zurlo..2020, Jorquera..21, Asensio-Torres..21,Huelamo..2022,Cugno..2023..NACO, Follette..2023}.  This low detection rate may be the result of high extinction at optical and near-infrared wavelengths due to the dust-rich nature of disk systems.   Nevertheless, upper limits in the planetary emission from several photometric bands have been set thanks to those surveys. In tandem, accurate constraints in the extinction from their host disk can provide key parameters to understanding the formation environments of protoplanets, and might be particularly useful for the design of future campaigns.

Current and upcoming facilities have devoted instruments for high-contrast imaging such as MagAO-X, ELT-METIS, and the Nancy Grace Roman Space Telescope for example. In addition to better instruments with deeper contrast limits, improved adaptive optics (AO) algorithms/techniques are also pushing the achievable contrast. Image processing techniques such as Reference Differential Imaging \citep{Grady..1999}, Angular Differential Imaging \citep{Marois..2006}, Polarization Differential Imaging \citep{Kuhn..2001}, and Spectral Differential Imaging \citep{Claudi..2008} attempt to make the best characterization possible of the stellar scattered light to subtract the stellar emission contamination from the planetary signal. Thanks to the capabilities of new instruments and state-of-the-art high-contrast imaging techniques, the detection limit will be pushed even beyond Jovian planets to the Super Earths, and much closer to their host stars, reaching the terrestrial planet-forming region. Thus, understanding the observational limitations and the expected extinction from circumstellar disk is  pivotal to the success of current and future protoplanet-hunting surveys.

Other approaches in the literature have tried to address the observability of accreting planets in circumstellar disks. \cite{Chen..22} show predictions for the observability of circumplanetary disks for the European Extremely Large Telescope, E-ELT,  and the  METIS instrument alongside JWST \citep[see also,][]{Boccaletti..22, Girard..22, Oberg..23}. \cite{Sanchis..2020} used 3D hydrodynamical simulations of massive planets to predict the extinction  by integrating the column density and converting it to A$_{\lambda}$ using ISM extinction curves. All these works show that a correct understanding of planetary emission and extinction will be key in exploiting the full potential of current and future facilities.

Multiple constraints on the dust content of protoplanetary disks have been inferred thanks to multiple observational surveys. The Atacama Large Millimeter Array (ALMA) has put limits to the number of millimeter grains in these disks and has characterized their spatial distribution alongside planetary architectures able to explain the disk millimeter structure, while infrared surveys have provided supplementary information about the $\mu$m-sized dust at the disk surface. Combining all this information becomes key to measuring the dust extinction in optical and infrared wavelengths, and to ascertain the observability of prototoplanets in typical growing conditions at those same wavelengths.

In this work, we focus on the predictions of the expected extinction for different geometries and grain sizes, which can be used with the predictions of the magnitudes of accreting planets \citep[e.g.][]{Zhu..2015} to correct for point source detections with multiple photometric points.   
We analyze the extinction dependencies with the inclination and the assumed maximum grain size of dust populations and planetary mass. We discuss what wavelengths and architectural configurations are best suited for planet-hunting campaigns given the expected contrast and angular separations. A careful selection of filters for a given target can optimize the requested time of observations maximizing the scientific output and characterization of the disk and protoplanet properties. 

The present work is organized as follows: we describe the methodology in Section \ref{sec:methods} and show the results in Sect. \ref{sec:results}. Then, we discuss the presented results in the current scenario for high-contrast imaging campaigns in Section \ref{sec:disc}. We summarize the main findings of our work in Section \ref{sec: summary}.
Additionally, in Appendix \ref{Appendix:disks} we report values of the extinction in protoplanetary disks that are commonly targeted in current high-contrast imaging campaigns as possible protoplanet-hosts disks.

 \section{Numerical Modeling and Methods} \label{sec:methods}

We explore the extinction in planet-forming disks under realistic conditions and different disk inclinations. We run hydrodynamical simulations using the \texttt{FARGO3D} code \citep{FARGO3D} with embedded protoplanets. We also calculate the changes  in the optical depth for different dust populations by varying the maximum grain size in the small grains, leading to differences in dust extinction.

\subsection{Hydrodynamical Simulation Setup}

With the goal of producing representative scenarios to calculate the extinction,  we employ the hydrodynamical code \texttt{FARGO3D} to produce synthetic density distributions of protoplanetary disks assuming different planetary  masses \citep{FARGO3D}. We ran 3D simulations in a  spherical grid with $(256,512,64)$ cells in radius, azimuth and colatitude respectively. The azimuthal coordinate spans the entire circumference from 0 to 2$\pi$ and the colatitude covers from 1.32 to 1.82 radians. The radial coordinate ranges from 0.4 to 4.5 in code units, which are scaled depending on the position of the embedded planet, which is located at 1.0 code units. Except the specific-disk dedicated models, an aspect ratio of 0.08 was assumed at the planet's separation with a flaring index of 1.15. We run the simulations for 1000 orbits when the disk has reached a steady state. We test different planet-to-star mass ratios, $q$, to reproduce the effect that different targeted gas giants can have. Thus, we use three different values for $q$, which are $q=5\times 10^{-3}$, $q=10^{-3}$, and $q=3\times 10^{-4}$. Assuming a Solar mass star, they correspond to a 5 M$_{\mathrm{Jup}}$, a 1 M$_{\mathrm{Jup}}$, and a 1 M$_{\mathrm{Sat}}$ planet respectively.

We use the multifluid capabilities of \texttt{FARGO3D} \citep{Benitez..2019} to also take into account the distribution of the solids in our models. The multifluid capabilities of \texttt{FARGO3D} follow the evolution of different dust populations with a given Stokes number, taking into account the dust differential velocity in the Epstein regime and the dust diffusion. The Stokes number in the Epstein regime can be written as \citep{Drazkowska..PPVII}:

\begin{equation}\label{Eq:ST}
    St = \frac{2}{\pi} \frac{a \rho_{\rm grain}}{\Sigma},
\end{equation}

with $\rho_{\rm grain}$ the dust grain density, $\Sigma$ the gas surface density and $a$ the grain size. To follow the dust distribution in the gap, we use the Stokes number corresponding to the different maximum grain sizes explored in our models assuming the surface density in the planetary orbit for each one of the different planets and an average grain density of $\rho_{\rm grain}=3.5$ g cm$^{-3}$. Thus, to match the same grain sizes at the center of the gap, the used Stokes numbers differ from model to model.

Our models assume a constant $\alpha$-viscosity prescription \citep{Shakura-Sunyaev} across the disk with $\alpha=10^{-3}$. The assumed $\alpha$ value can have significant impacts on the level of extinction as it is a key parameter in determining the local level of depletion inside planet-carved gaps \citep{Kanagawa..2015,Kanagawa..2017} and the dust settling \citep{Fromang..2009, Johansen..2014}. We discuss possible caveats of the modeling further in Section \ref{sec:disc}.

Additionally, we report specific extinction maps of planet-hosting disk candidates in Appendix \ref{Appendix:disks}. These specific disks have confirmed detected protoplanets or they have planned high-contrast imaging planet-hunting observations with several instruments, which makes them objects of interest for planet formation research.

\subsection{Observational Constraints}

Many surveys and observational programs have put constraints on the absolute and relative amount of small dust, i.e., $\mu$m-sized dust and millimetric grains (pebbles) through different methods. However, their relative mass fractions cover a wide range of values.  Dust growth in protoplanetary disks shifts the expected extinction compared to the ISM, which is mostly composed of $\mu$m-size grains \citep{Mathis..1977, Birnstiel..2018}. Thus, while dust growth leads to a slight decrease of extinction in optical wavelengths, it may lead to an increase of the opacity in the near- and mid-infrared with respect to the ISM due to a larger share of sub-millimeter grains, although that increase is less significant. The separation between two populations of small ($0.1-1~\mu$m) and large ($1~\mu$m~$-$ 1 mm) grains is supported by the dust settling. Different sizes have distinct drag interaction regimes with the gas in the disk. Smaller grains will be more coupled to the gas, while larger millimetric grains will quickly settle to the midplane \citep{Barriere..2005, Dullemond..Dominik}. This effect has been confirmed by observations of edge-on disks with ALMA \citep{Villenave..2020}, showing clear razor-thin distributions and dust size stratification, where the larger millimeter grains are concentrated in the disk midplane, and from scattered-light images showing the $\mu$m-sized grains in the surface \citep{Avenhaus..2018}. This vertical distribution is distinctive of the dust-settling processes. While for the sake of simplicity models adopt two populations, it is generally assumed they follow a continuous MRN-like distribution \citep{Mathis..1977}.

The surface density of large mm-sized grains in protoplanetary disks is constrained via high-resolution observations with ALMA which have improved our understanding of the dust distribution \citep{DSHARP..I, MAPS_XIV}. Thus, the mass and distribution of large grains is fairly well-characterized, and hence, their contribution to the extinction. Nevertheless, different disks will have different surface densities depending  on their unique features. Based on the formalism of \citet{Shakura-Sunyaev}, \citet{Lynden-Bell..1974}, and \citet{Hartmann.98}, the expected characteristic surface density, $\Sigma_c$, at the characteristic radius, $R_c$, for a viscously-evolved disk is:

\begin{equation}\label{Eq: sigma c}
    \Sigma_c = \frac{(2-\gamma)M_{\mathrm{disk}}}{2\pi R_c^{\gamma}(R_{\mathrm{max}}^{2-\gamma}-R_{\mathrm{min}}^{2-\gamma})},
\end{equation}

\noindent where $R_{\mathrm{min}}$ and $R_{\mathrm{max}}$ are the inner and outer disk radius respectively, M$_{\mathrm{disk}}$ the disk mass and $\gamma$ the exponential factor of the radial power-law decay. \citet{Tazzari..et..al..2016} and \citet{ MAPS..V..Coco} find that values of $\gamma \sim 1$ produce reasonable fits to the dust and gas distribution of several protoplanetary disks. Thus, for simplicity, we assume that $\gamma=1$ for our simulations. Equation \ref{Eq: sigma c} shows that the characteristic surface density scales with disk mass and is dependent on the disk size as well.  Typical dust surface densities span between 10$^{-4}$ to 10 g cm$^{-2}$ depending on the overall dust mass of the disk and the location within it. In the sources observed by the ALMA Large Program ``Molecules with ALMA at Planet-forming Scales (MAPS)'', \cite{MAPS..V..Coco} report that the millimeter grains dust surface densities at the characteristic radii (80-200 au) range from 0.05 to 0.28 g cm$^{-2}$. For comparison, in the initial conditions of our base model, we assume a planet is orbiting at $R_p$=100 au, a dust disk mass of 10$^{-4}$ M$_{\odot}$, $R_{\mathrm{max}}-R_{\mathrm{min}}$ = 410 au and M$_{\mathrm{disk}} = 0.01$ M$_{\odot} = 10$ M$_{\mathrm{Jup}}$. Using Equation \ref{Eq: sigma c}, the base characteristic dust surface density is $\Sigma_c=$ 0.003 g cm$^{-2}$. For our base model, the disk mass is 0.01 M$_{\odot}$, which is lower than typically observed disks in high resolution. Nevertheless, this disk mass value is probably more representative of the typical protoplanetary disk mass in known start-forming regions than one of the disks with resolved dust substructures  \citep{Bae..2023, Nienke..Pinilla..2023}. Appendix \ref{Appendix:disks} lists extinction values from our models for a subset of disks with suggestive evidence of the presence of protoplanets.

In the case of smaller grains, thermochemical modeling has attempted to constrain the amount of small grains that enable the formation of certain key chemical tracers. \citet{Bosman..21} and \citet{ Calahan..2023} find that the amount of small dust influences the observed chemical composition (traced by C$_2$H, CH$_3$CN, and HC$_3$N) as they are key players in the thermochemical balance of the disk. To match resolved ALMA gas phase emission of these species with detailed thermochemical models requires reduced (0.1\% by mass) contributions of small grains to the dust surface density in a couple of disks, HD 163296 and MWC 480, and 25\% in the case of TW Hya \citep{Calahan..2023}. 

 Additional studies shed light on the amount and distribution of small grains by observing the scattered light emission with SPHERE \citep{Van..Boekel..2017, Avenhaus..2018}; from which \cite{Van..Boekel..2017} required ~0.1-5\% of small dust content with respect to large grains to reproduce the scattered light emission of the TW Hya disk. The analysis of polarization data provides another constraint in tandem with millimeter emission\citep{Ohashi..Kataoka..2019, Rich..2021}. One unique extinction measurement comes from  MUSE observations of the PDS 70 system by taking the ratio of the emission of hydrogen lines. For example, \cite{Hashimoto..et..al.2020} using an ISM extinction law give estimates of A$_{H\alpha}>$2.0 and $>$1.1 mag for PDS 70b and PDS 70c respectively.  However, there is a degeneracy between the amount of small dust and its maximum grain size that can replicate such values. Hence, the small grain content is still uncertain and we treat it as a variable, exploring grain fractions from 10\% to below 1\% (by total dust mass) based on the estimates mentioned above. Nonetheless, if constraints are present in the literature, we used them for the disk-specific models in Appendix \ref{Appendix:disks}.

\subsection{Extinction Calculation}

\begin{figure}
    \centering
    \includegraphics[width=0.99\linewidth]{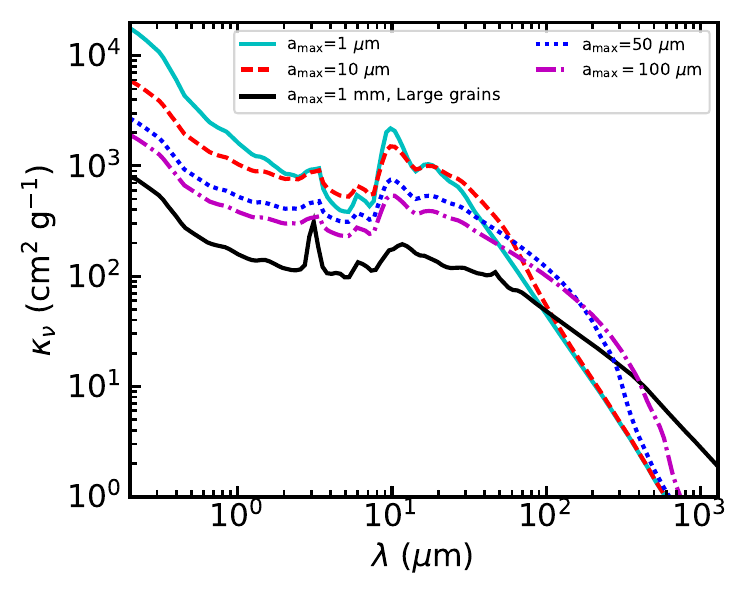}
    \caption{Absorption curves assuming different maximum grain sizes for the population of small dust and the large dust showing their wavelength dependency.}
    \label{fig:kappa}
\end{figure}

The opacity in the disk is mostly dominated by dust \citep{Hensley..23}, so our extinction estimations are derived from dust opacity values. We used two dust populations for the disk radiative transfer modeling to account for the small grain and large grains contribution to extinction. The first population holds 99\% of the disk dust mass and corresponds to large grains. The second population corresponds to the small grains holding 1\% of the disk dust mass. We also compute models with a 90/10\% share of the large to small dust grains distributions.  However, from the opacity values we show in Figure \ref{fig:kappa}, below the 1\% level, the optical depth contribution of the small grains becomes negligible compared to the  the large grains at infrared wavelengths. We use the same dust mixture for both populations, which is  40\% in organics, 33\% astrosilicates, $\sim$ 8.8\% in troilites, and a water ice mass fraction of 20\%. The opacity constants used to calculate the dust opacity of the mixture are taken from \cite{Henning..Stognienko..1996} for organics and troilite, \cite{Draine..2003} for the astrosilicates, and from \cite{Warren..Brandt..2008} for water ice. We set a constant dust-to-gas mass ratio across the disk of 0.01 to set the dust mass \citep{Sandstrom..2013}, although this parameter can be scalable to match the desired target.  

To include the contribution of different dust grain sizes to the dust population we use the standard dust size distribution following the \cite{Mathis..1977} power law for both dust populations: 

\begin{equation}
    n_0(a) \propto a^{-p}da,
\end{equation}

\noindent with $a$ being the grain size, and $p$ the power law index, which we set to 3.5. The large grains population ranges from $a_{\rm min}$ = 0.05 $\mu$m to $a_{\rm max}$ = 2.5 mm, which is within the maximum grains sizes inferred by the MAPS survey, showing that the sources observed by the MAPS collaboration have dust grains present up to 3 mm \citep{MAPS_XIV}. The small dust population has a fixed minimum grain size, $a_{\rm min}$ = 0.05 $\mu$m, while the maximum grain size is variable depending on the model. The opacity for a given dust population will be given by its composition and also by the grain sizes within the population itself, i.e., the maximum grain size causes variation in the shape of the opacity curve, which can produce differences in the expected extinction. Thus, we explored four different maximum grain sizes for the small dust, which are $a_{\rm max}$ = 1; 10; 50, and 100  $\mu$m.  The different opacity curves as a function of wavelength are illustrated in Figure \ref{fig:kappa}.

We then use \texttt{RADMC-3D} \citep{RADMC-3D} to calculate the optical depth of the disks at different lines/bands of interest which are listed in Table \ref{Tab:Filters}. These lines/bands correspond to typical accretion tracers and the filters of the state-of-the-art or 30-meter planned telescopes such as the VLT, JWST or the ELT. These upcoming facilities will have more powerful instruments to probe milliarcsec resolution with deeper sensitivity \citep{Benisty..22} while having  strong synergy with the reported wavelengths in this work.

The extinction is then calculated from the optical depth images at varying inclinations starting from a face-on orientation of  0\degg \ until reaching 60\degg\ in constant increments of 10\degg . For our extinction retrieval we assume that the planet is located along the minor axis of the disk except for the specific disk cases described in Appendix \ref{Appendix:disks}. Possible variations of the circumstellar extinction depending on the position angle of the planet on its orbit are further explored in Appendix \ref{Appendix:PA_devs}. In Figure \ref{fig:op_angle} we illustrate the line of sight overlap with the total dust density in the disk for different viewing angles at the location of the planet. We do not attempt inclinations beyond 60\degg\ as the extinction starts to increase quickly due to the bigger overlap with the circumstellar disk, and the role of the planet's influence on the host disk becomes less dominant. Thus, at these inclinations the extinction becomes dominated by the circumstellar disk itself outside the planet-carved gap. The sightlines towards highly inclined disks ($i>$60\degg) captures higher magnitudes of disk extinction; this will hamper the success of planet-hunting campaigns. For intermediate inclinations, even though the inclination increases, carefully chosen targets can minimize extinction, improving the chances of success in finding protoplanets.

We created synthetic images using \texttt{RADMC-3D} \citep{RADMC-3D} to recover the optical depth at the wavelengths of interest from our hydrodynamical models.  We take advantage of the relationship between the optical depth, $\tau_{\lambda}$, and the extinction:

\begin{equation}
    A_{\lambda} = 2.5\log{(e)}\tau_{\lambda}= 1.086\tau_{\lambda},
\end{equation}

\noindent to calculate the expected extinction at the location of the planet for different disk inclinations, planetary masses, and the grain size of the small grains.

\begin{figure}
    \centering
    \includegraphics[width=0.99\linewidth]{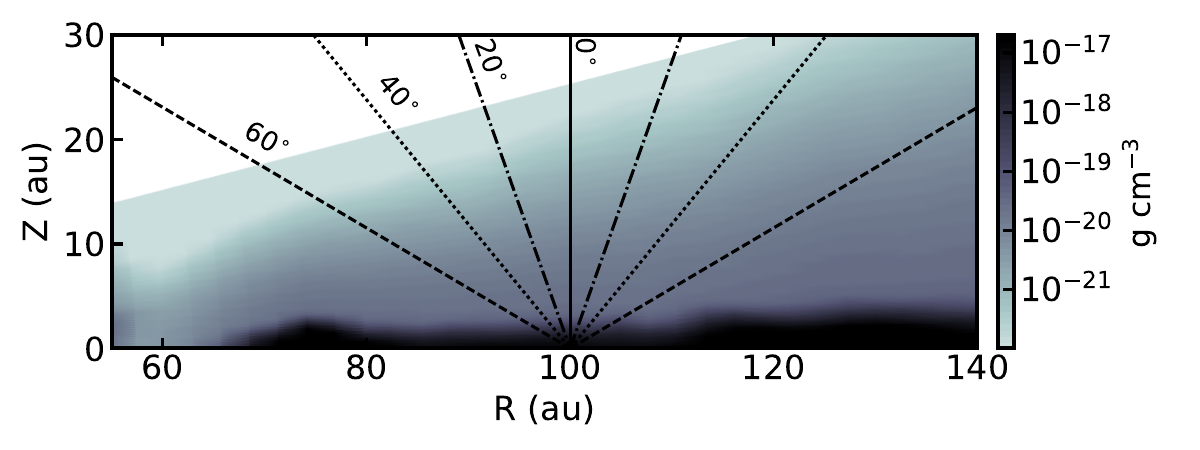}
    \includegraphics[width=0.99\linewidth]{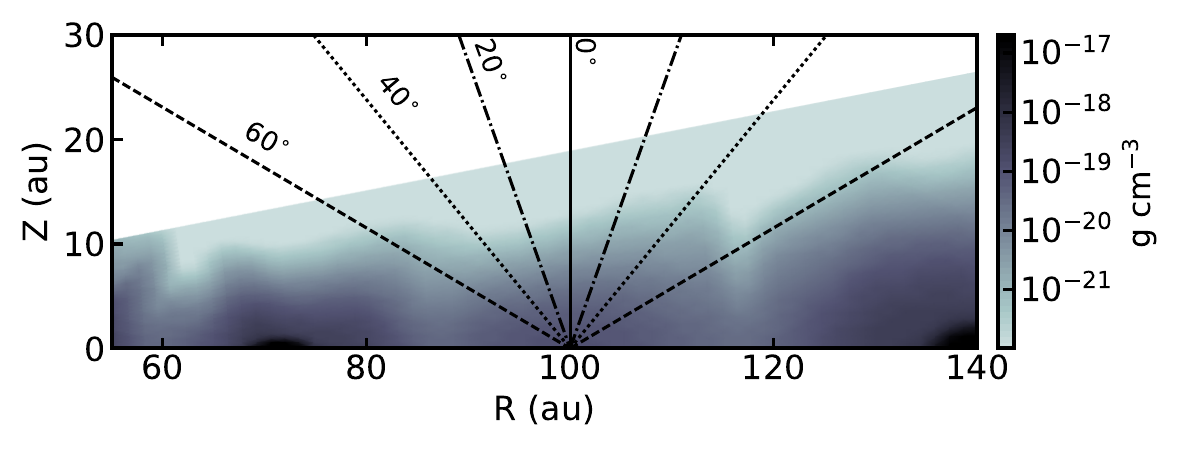}
    \includegraphics[width=0.99\linewidth]{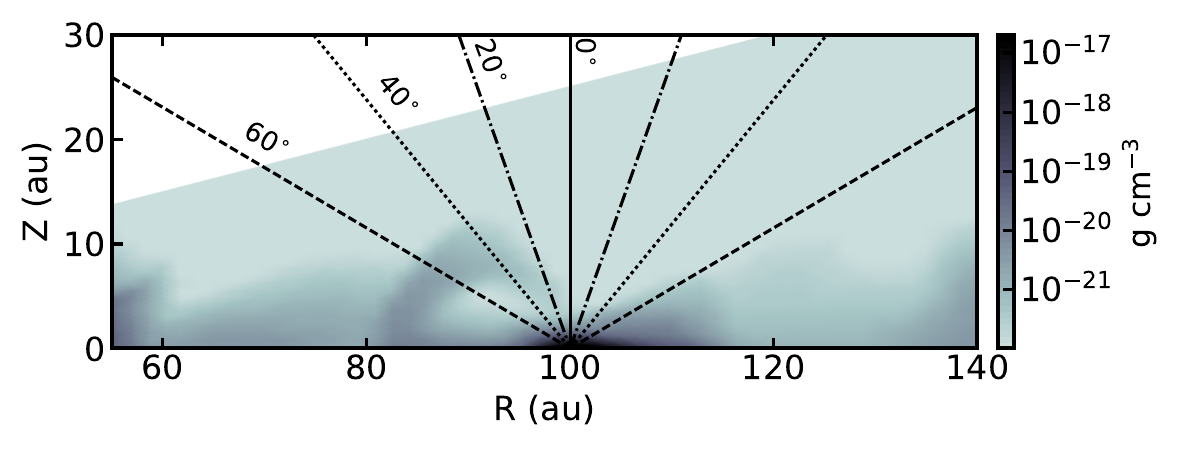}
    \caption{Plane cut of the dust density at the planetary azimuth of the M$_{\rm Sat}$, M$_{\rm Jup}$ and 5M$_{\rm Jup}$ planets from top to bottom. The different lines illustrate a few of the different viewing angles explored in this work from 0$^{\circ}$ to 60$^{\circ}$.}
    \label{fig:op_angle}
\end{figure}

 \begin{table}
 \caption{Selected bands/lines for the extinction calculations}
\begin{center}
\begin{tabular}{cc|cc}
\hline
\hline
Band & Representative & Line & Representative  \\
 & Wavelength & & Wavelength  \\
\hline
H &  1.6 $\mu$m & H$\alpha$ & 0.656 $\mu$m \\
K &  2.2 $\mu$m  &  Pa$\alpha$ &  1.88 $\mu$m  \\
L/L' &  3.5/3.79 $\mu$m  & Br$\alpha$ &  4.05 $\mu$m \\
M' &  4.8 $\mu$m & Br$\beta$ & 2.63 $\mu$m\\
N1 &  8.65 $\mu$m  & & \\
N2/F1140M &  11.2/11.4 $\mu$m  & & \\
\hline 
\end{tabular}
\end{center}
\label{Tab:Filters}
\end{table}

\section{Results}\label{sec:results}

\begin{figure*}[h!]
    \centering
    \includegraphics[width=0.99\linewidth]{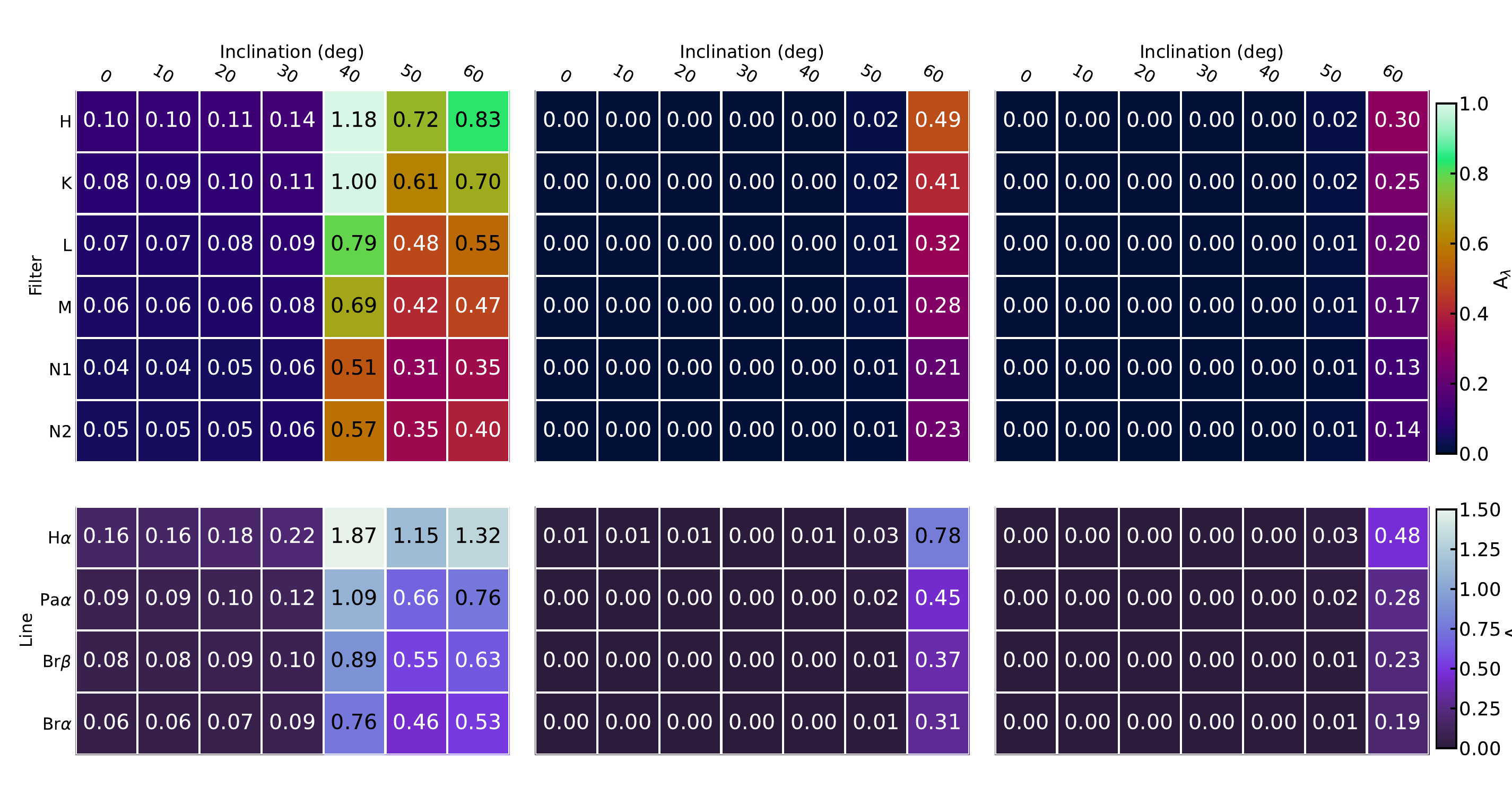}
    \caption{Measured extinction for different planet-star mass ratios, $q=3\times 10^{-4},10^{-3},5\times 10^{-3}$ for left, center and right panels respectively, and disk inclinations assuming a maximum grain size for the small grain dust population of 10 $\mu$m.}
    \label{fig:10 microns}
\end{figure*}

\begin{figure*}
    \centering
    \includegraphics[width=0.99\linewidth]{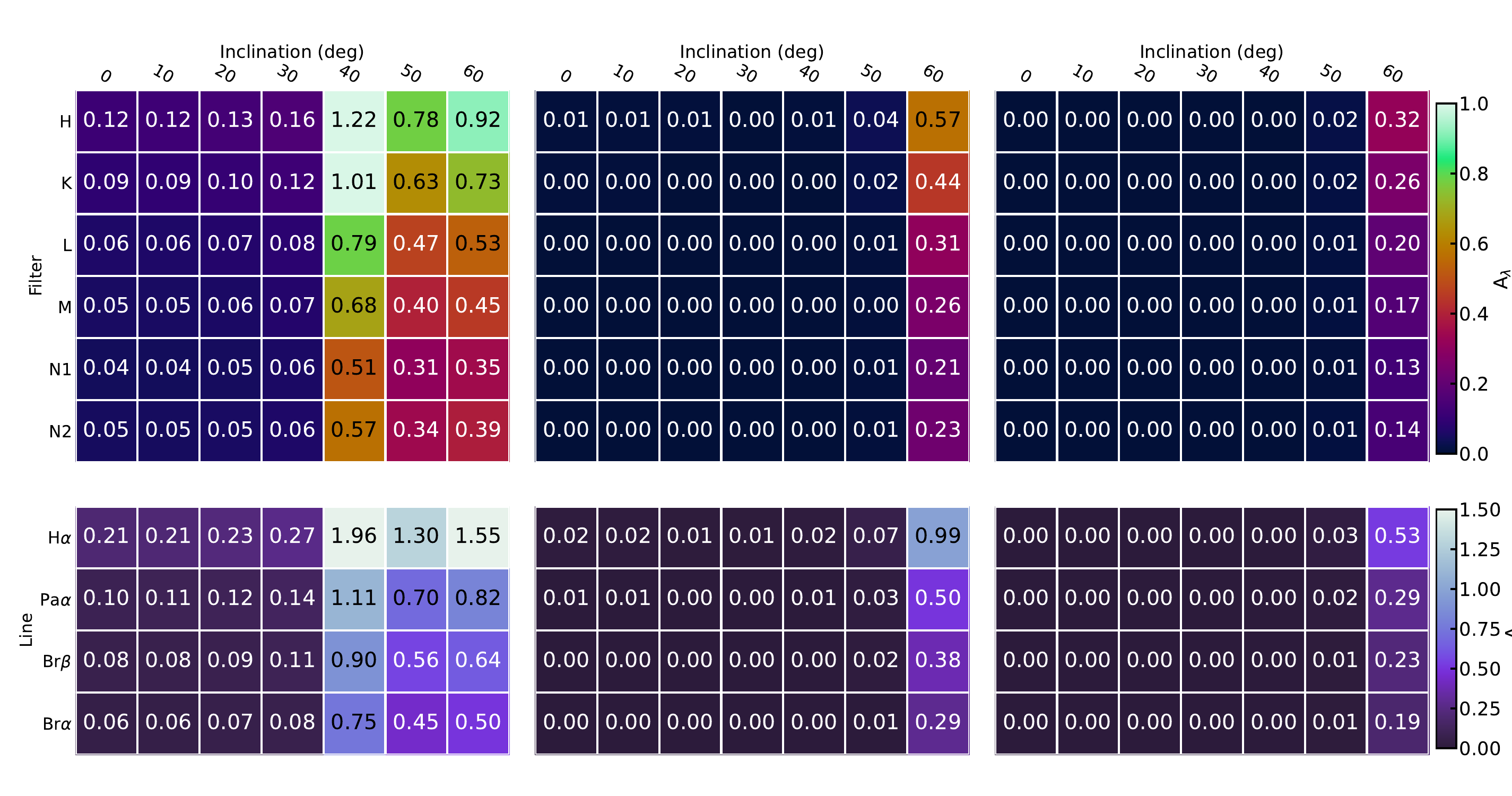}
    \caption{Same as in Figure \ref{fig:10 microns} with a maximum grain size of 1 $\mu$m for the small grains.}
    \label{fig:1 microns}
\end{figure*}

\begin{figure*}
    \centering
    \includegraphics[width=0.99\linewidth]{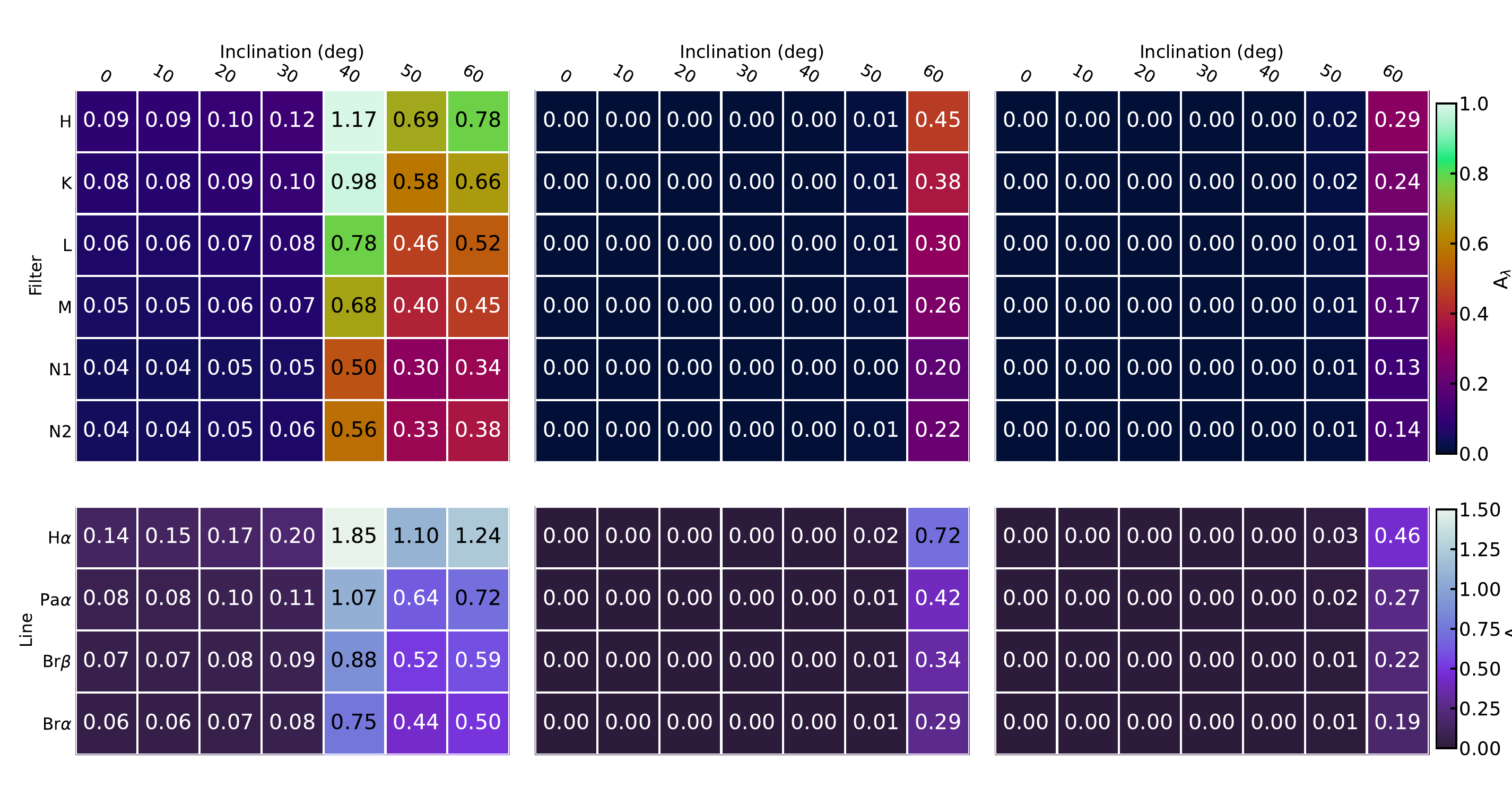}
    \caption{Same as in Figure \ref{fig:10 microns} with a maximum grain size of 50 $\mu$m for the small grains.}
    \label{fig:50 microns}
\end{figure*}

\begin{figure*}
    \centering
    \includegraphics[width=0.99\linewidth]{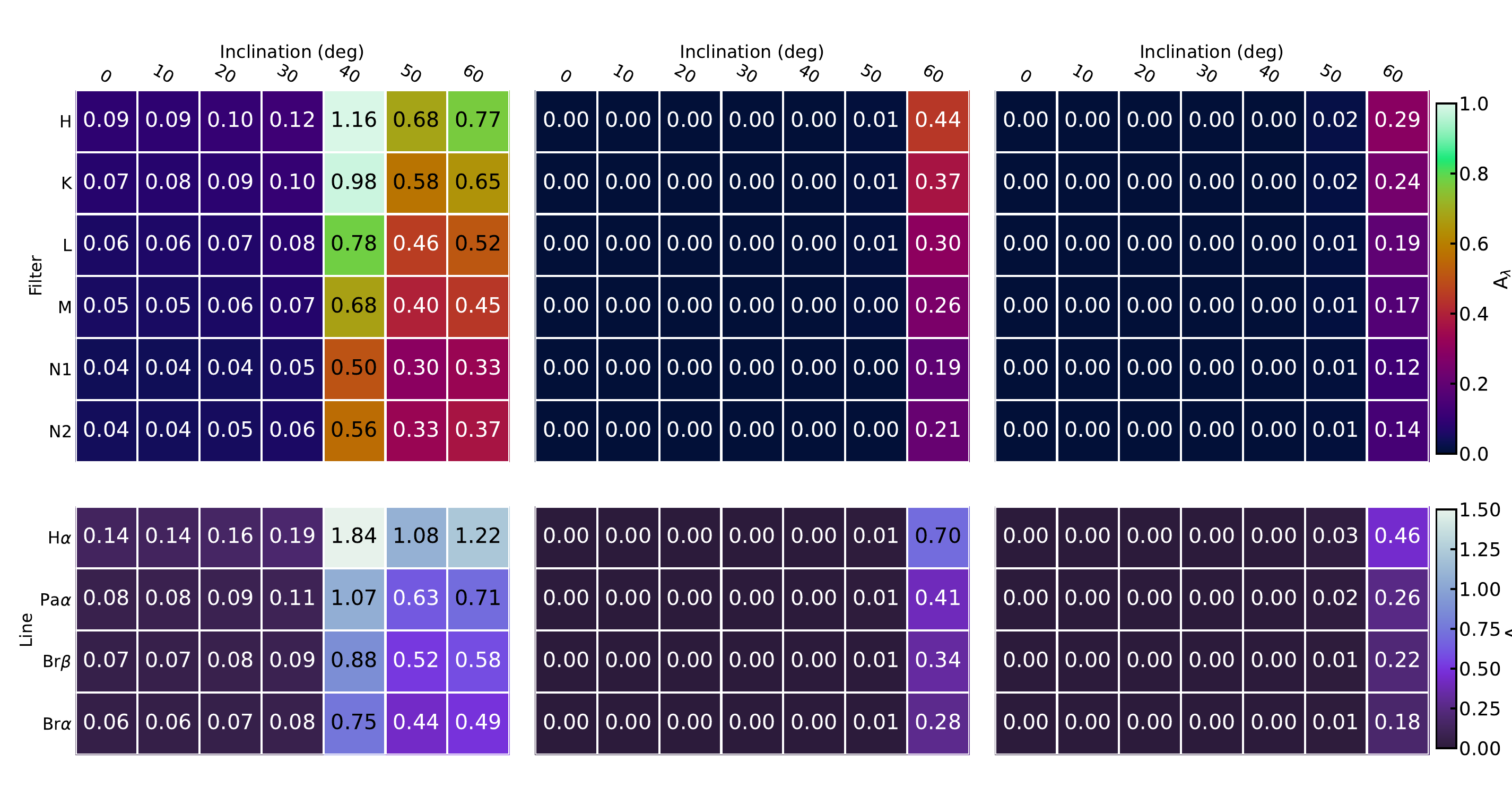}
    \caption{Same as in Figure \ref{fig:10 microns} with a maximum grain size of 100 $\mu$m for the small grains.}
    \label{fig:100 microns}
\end{figure*}

\subsection{Planet Mass}

We show the different extinction values for the three different planetary masses in Figure \ref{fig:10 microns}. If we compare the same filters and the same inclinations, the extinction decreases as the planet mass increases. The trend is expected as more massive planets produce deeper gaps \citep{Kanagawa..2015}, which reduces the column producing extinction along the line of sight. From the tables, it is noteworthy that in the case of a massive Jovian planet, 5 M$_{\mathrm{Jup}}$, the extinctions stay moderately low at significantly inclined disks, which is related to its bigger sphere of influence on the disk. It is also noteworthy that the width of the gaps increases as the planetary mass increases. Therefore, the more massive planets have  less overlap between the circumstellar disk and/or gap edges with the line of sight for moderate inclinations.

The same pattern of decreasing  extinction with planetary mass is recovered in Figures \ref{fig:10 microns}, \ref{fig:1 microns}, \ref{fig:50 microns} and \ref{fig:100 microns} assuming different dust populations as the trend depends on the integrated column density along the line of sight and not in the opacity unless the composition of the dust grains changes drastically for different planetary masses, which is beyond the scope of this paper.

\subsection{Inclination}

For very inclined disks, $i>60$\degg, the line of sight has a larger overlap with the circumstellar disk beyond the gap width tracing regions that are not being depleted. Hence, the extinction is quite large.  However, assuming a planetary mass of 5 M$_{\mathrm{Jup}}$ the planet still depletes the disk material making the extinction lower than its counterparts with lower masses. For close to face-on values, we would be looking straight on top of the planet, with much smaller overlaps with the circumstellar disk and much more with the sphere of influence of the planet. As more massive planets produce deeper depletions inside the gap, the extinction decreases accordingly.

For intermediate inclinations, 60\degg$>i>20$\degg, the line of sights starts to go through the gap carved by the planet producing lower extinctions. Nevertheless, the trend in inclinations will be more dependent on the specific parameters of the disk. The tables in Figs. \ref{fig:10 microns}, \ref{fig:1 microns}, \ref{fig:50 microns}, \ref{fig:100 microns} show that the Saturn mass planet has a peak at 50\degg, while the Jupiter mass planet at  60\degg. This effect is caused by the pile-up of dust particles at the edge of the gap, which is a strong dust trap, so it produces a big sharp enhancement in the extinction towards the disk. The angle at which this sharp change occurs will depend on the gap width and the strength of the dust trap.

We also remark that the lowest extinctions are not produced for face-on disks ($i=0$ \degg ), but for slightly inclined disks, although the difference is not substantial. We link this phenomenon to the asymmetries in the density fields around the planet, i.e., there is a slightly larger column density right on top of the planet in a narrow polar region where the gas is being funneled towards the planet \citep{Fung..et..al..2015, Sanchis..2020}. Even if the millimeter disks are very thin (geometrically), massive planets can cause local dust stirring, i.e., they still push dust from the midplane \citep{Binkert..et..al..21}. The same funneling region is bypassed when the disk is observed slightly tilted at a low inclination, avoiding the bulk of the disk extinction and the polar funneling on top of the planet. The width of the funneling effect decreases with increasing planet mass until it becomes negligible for super Jovian masses.  

\subsection{Grain Size Dependency}

As observed from the opacity curves from Figure \ref{fig:kappa}, the dust opacity behavior is not linear between different grain sizes, however, with multiple photometric points it can be disentangled. Overall, from 1 to 30 $\mu$m, the lowest opacity is reached by 50 $\mu$m grains, while the largest is in general for the 10 $\mu$m maximum grain size. As the assumed grain size is smaller, the behavior becomes different reaching a minimum between the near- and mid-infrared, between 5 and 10 $\mu$m, and then increases towards the far-infrared before starting to drop in the submillimeter wavelengths. Regardless of the explored grain sizes, the minimum extinction is reached around 8.6 $\mu$m in the mid-infrared. However, from Figures \ref{fig:1 microns}, \ref{fig:50 microns} and \ref{fig:100 microns} the overall filter and line dependency changes, which provides the opportunity to constrain the small dust content around protoplanets by observing them in multiple filters, even if the protoplanet is not observed in all of them.

\subsection{Extinction curve in infrared wavelengths}

\begin{figure}[h!]
    \centering
    \includegraphics[width=0.99\linewidth]{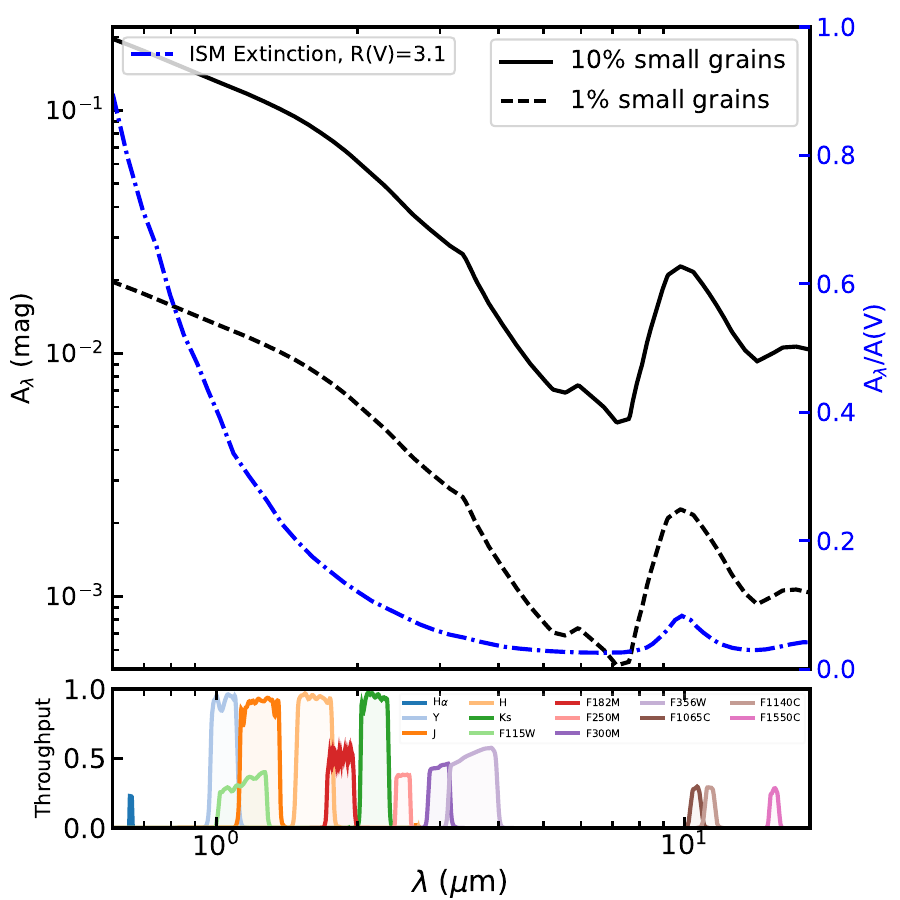}
    \caption{Extinction curve for a face-on disk with a Jupiter mass planet overlapped with the bandpasses of a few coronographic filters of NIRCAM, MIRI, and SPHERE. For different amounts of small grains content, there is a different wavelength dependency  of the extinction. For small grains content below 1\%, the opacity in infrared wavelength starts to be dominated by the millimeter grains. Multiple photometric points can hel constrainin the extinction curve and hence the small grains mass share in disks. The blue dash-dotted line shows the extinction for typical diffuse ISM with R(V)=3.1 for comparison \citep{Gordon..23}.}
    \label{fig:filters}
\end{figure}

We illustrate the expected extinction curves in Figure \ref{fig:filters} for our model with a Jupiter mass planet orbiting at 100 au in a face-on disk overlay with the throughput of multiple coronographic filters in the NIRCAM, MIRI and SPHERE instruments + the H$\alpha$ filter as a tracer of gas accretion. In the Figure, we compare the extinction contribution assuming different contributions of small grains to the total dust mass in the disk. Even in the absolute absence of small grains, there is a "floor" expected extinction towards the midplane, which is set by the millimeter grains. For comparison, we also add the extinction curve for diffuse ISM, R(V)=3.1 \citep{Gordon..23}, showing that larger dust grains in the protoplanetary disk increase the dust extinction in the mid-infrared with respect to the ISM.  However, even in this case, the extinction at longer near-infrared and mid-infrared wavelengths is still lower than the one expected in the optical. As the blackbody emission of a massive planet and the one of a CPD are expected to emit mostly at mid-infrared wavelengths \citep{Zhu..2015}, the extinction being lower helps since the contrast is also more favorable at these wavelengths.

The extinction increases quickly from the mid-infrared to optical wavelengths, hardening the success rate of planet-hunting observational campaigns in that wavelength regime where the expected contrasts with the star are already not optimal if they are not carefully designed or the targeted sources are too extinct. However, low-mass disks with planets in wide-orbit separations can have extinctions of $\sim$2 mag, even for optical wavelengths or for the H$\alpha$ line, which can still be achieved with current instrumental capabilities. Detailed elaborated surveys can then lead to the detection of protoplanets. Moreover, the use of multiple photometric points in already observed sources helps in understanding the properties of the dust grains and their growth in planet-forming environments.

\subsection{Scalability of the extinction}

If we only consider the gas evolution and a constant dust-to-gas ratio across the disk, our simulations are unit-free, i.e. they can be scaled attempting to find a general relationship to estimate the dust extinction towards embedded planets in protoplanetary disks. The measured optical depth is proportional to the integrated column density along the line of sight, which is scalable with the separation of the planet, a given disk mass, and disk size. Given the linear relationship between optical depth and extinction from Equation \ref{Eq:scaling}. Therefore, the expected extinction for a given planet location, $R_\mathrm{p}$, $\zeta$ dust-to-gas ratio, and disk mass, $M_{\rm disk}$, is:

\begin{footnotesize}
\begin{equation}\label{Eq:scaling}
    A_{\lambda}(\zeta, M_{\rm disk}, R_\mathrm{p}) \propto \zeta \Big(\frac{M_\mathrm{disk}}{0.01\ \mathrm{M}_{\odot}} \Big)\Big(\frac{R_\mathrm{p}}{100\ \mathrm{au}}\Big)^{-2} \Big(\frac{R_{\mathrm{out}}-R_{\mathrm{in}} }{450 - 40 \ \mathrm{au}} \Big)^{-1}. 
\end{equation}
\end{footnotesize}

\noindent Equation \ref{Eq:scaling} also includes $R_{\mathrm{out}}$ and $R_{\mathrm{in}}$ which represent the outer and inner radius of the disk, i.e., the disk size. In our approach, the scaling is linear with the inverse of the disk size given that in our model the initial surface density scales with $r^{-1}$, although the general dependency is given by $r^{\gamma-2}$.

\begin{figure}
    \centering
    \includegraphics[width=0.99\linewidth]{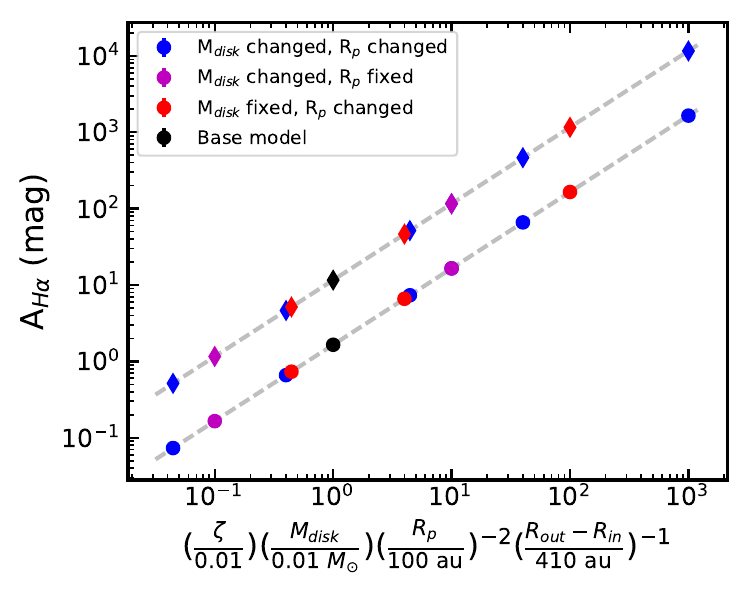}
    \caption{Scaling of the extinction  at H$\alpha$ for the hydrodynamical simulation with an embedded Jupiter, or q=$10^{-3}$, for a face-on disk (circles) and one with an inclination of 60\degg (diamonds). The measured extinctions follow the dependency on the inverse of the square of the planetary separation and linearly with respect to the total amount of dust mass in the disk. Moreover, the offset between the different points of view roughly matches the expected factor due to inclination, i.e.,  $\sec$ (60 \degg ).}
    \label{fig:scaling}
\end{figure}

The relationship was tested by calculating the optical depth with \texttt{RADMC-3D} in different scalings of the hydrodynamical output varying the disk mass and the location of the planet while keeping the units-free domain of the simulations constant. In Figure  \ref{fig:scaling} we show an example of the extinction scaling of the H$\alpha$ line with the disk mass, the inverse of the square of the Jovian planet separation, and the disk size. As the disk size increases the disk mass is distributed in a larger area, which decreases the absolute values of the surface density and then the extinction. Planets at wider orbital separations will suffer from less extinction from circumstellar material due to the radial decay of the surface density.

It is relevant to mention that for different disk masses, and thus, surface densities, the Stokes numbers are different. From Equation \ref{Eq:ST}, $St \propto \Sigma^{-1}$, meaning that keeping the disk size fixed, decreasing the disk mass by an order of magnitude increases the Stokes number of the same particle size by an order of magnitude. Therefore, the radial drift of dust particles of the same size is expected to be faster in less massive disks, effectively depleting them in shorter timescales and decreasing extinction. The opposite trend occurs for massive disks where the dust is allowed to grow to larger sizes with a less efficient radial drift. We expect that the curve in Figure \ref{fig:scaling} acts as an upper limit with the actual values close to the ones expected in the curve in the top-right corner where dust coupling occurs in shorter timescales. For the disks in the bottom-left regime, the expected extinction should be lower since radial drift is faster for larger grains, reinforcing our statement that less massive disks should suffer from less significant effects.

\section{Discussion}\label{sec:disc}

We complement the work done by \cite{Sanchis..2020} exploring different inclination angles and assuming dust growth, i.e., larger grains than in the ISM. Additionally,  we present predictions for different grain size populations calculating their opacity curves for a mixture of different compositions replicating the one done by \cite{Birnstiel..2018} in the DSHARP survey.

\subsection{Breaking the Degeneracy and Retrieving Protoplanet Information}

From the explored extinction maps, we observe that depending on the location of a putative planet candidate, extinction may sink the probability of an observing run  in planet-catching observations. Nevertheless, our same modeling shows that for planets formed in the outer regions of protoplanetary disks, the extinction for most photometric bands decreases to almost negligible values.

Therefore, even though, the detectability of sub-Jovian protoplanets becomes a complicated task given the emission extinction from the circumstellar disk, a cleverly designed sample of target sources for planet-hunting campaigns should lead to their detection. So far, most campaigns target sources with detected substructures of protoplanet candidates from millimeter observations. However, those sources are biased towards bright protoplanetary disks in millimeter emission, i.e., they have very large optical depth and extinction values in infrared and optical wavelengths. Typically known protoplanet disks rich in dust substructure are more massive than the average \citep{Bae..2023, Nienke..Pinilla..2023}. For more compact sources in dust continuum emission with large gaseous disks, the dust extinction should be much lower than their dust-extended massive counterparts due to a lower dust column along the line of sight and a faster dust radial drift. With extinctions lower than 1 mag in most infrared filters for Jovian planets in a standard 0.05 M$_{\odot}$ disk and $<$2 mag for hydrogen lines.

Doing a characterization of detected protoplanets through multiple photometric filters, will not only allow us to characterize the extinction. By having multiple data points, not only the dust content and size distribution around the protoplanet but also its accretion rate and intrinsic emission can be constrained; and then possibly its formation pathway. Furthermore, as more protoplanets are characterized and detected, the measurement of their extinction can provide evidence about the geometrical nature of protoplanetary accretion, as different accretion mechanisms produce different extinctions due to the differences in the accretion streams \citep{Marleau..22}.

\subsection{Caveats}

It is noteworthy that our models assume a constant $\alpha$-viscosity prescription \citep{Shakura-Sunyaev} across the disk with $\alpha=10^{-3}$. The assumed $\alpha$ value can have significant impacts on the level of extinction as it is a key parameter in determining the local level of depletion inside planet-carved gaps \citep{Kanagawa..2015, Kanagawa..2017} and the dust settling \citep{Fromang..2009, Johansen..2014}. \cite{Kanagawa..2015} show that the level of depletion has a dependence on $\alpha^{-1}$. When the viscosity is high, the gap gets replenished by the disk in short timescales which leads to a higher extinction from the circumstellar disk and low depletions. The opposite happens for lower viscosity values, producing much more depleted gaps with geometrically thinner disks, which can lead to an overestimation of the extinction in these disks. Nevertheless, \citet{Rosotti..23} and \citet{Pizzati..2023} (and references therein) report that typically $\alpha$-values are $<10^{-2}$, with most disks actually having $\alpha$ values ranging between $10^{-4}$ and $10^{-3}$. 

Another parameter that has a significant influence on the expected disk extinction is the aspect ratio. For our runs, we assumed that the aspect ratio is 0.08 at 100 au, which is a typical value for protoplanetary disks. The thinner disks will suffer from more severe depletion due to the planet and also less overlap between the line of sight and the disk for inclined disks, while geometrically thicker disks will be less depleted with higher column density along the line of sight increasing their extinction and making them less desirable for high-contrast imaging surveys. Special consideration must be taken with disks with multiple planets where they can affect the gap depth by interacting with each other. Nevertheless, in the case that the separation between planets is big enough, the scaling provided in this work can give a good constraint on the expected extinction, and then on the small dust content. Farther, small grains are expected to be more coupled to the gas, while millimeter grains will settle towards the midplane creating a razor-thin structure \citep{Barriere..2005, Dullemond..Dominik, Villenave..2020}. Nevertheless, for face-on disks or disks with low inclinations, the contribution from both components is taken into account, while for high-inclination disks, the inclination itself increases the column density along the line of sight, abruptly raising the extinction, making those sources much less qualified for direct imaging campaigns.

The composition of the dust grains assumed they have water ice content. Thus, the scaling and estimated extinction values should only be used beyond the expected location of the water ice line, which should be located closer than 10 au for most low-mass and intermediate-mass stars with the exception of outbursting sources. Inside the water snow line, the transition between different sublimation points, which are key to calculating the opacity, takes place in smaller spatial scales, at which point it becomes harder to make reliable estimates of the extinction values without a source-specific model. In addition, we did not take into account the porosity of the grains, which can significantly reduce their opacity even by orders of magnitude, hence, decreasing the disk extinction as well \citep{Kataoka..2014, Miotello..2023}.

In our models with inclination, we assumed that the planet is located along the minor axis of the disk. For inclined disks, the extinction for different position angles will be different due to the disk geometry and the overlap of material along the line of sight. Even though in highly-inclined disk the circumstellar disk contribution to the extinction may be larger, the overall standard deviation covering all the protoplanetary orbit is less than 35\% in all cases, and much lower for low-inclination disks, with a standard deviation of the order of 10\% (see Appendix \ref{Appendix:PA_devs}).

\section{Summary}\label{sec: summary}

We present the extinction values inside planet-carved gaps from hydrodynamical simulations of multiple spectral bands of interest in high-contrast imaging planet-hunting campaigns in the infrared and optical.

We observed that the planetary mass and the mass of the host disk play a major role in the level of extinction along the line of sight towards a planet-carved gap among other observability constraints such as disk inclination and planet separation. Compared to the ISM extinction, larger grains present in protoplanetary disks increase the extinction in near- and mid-infrared wavelengths precisely where the emission of protoplanets may peak.

As expected, less massive disks can push the limit at which a protoplanetary disk may not suffer significant extinction, keeping it within the detectability threshold for protoplanets with large separations. There are multiple constraints that can make the detectability of an accreting planet difficult; however, given the same constraints, all the protoplanet detections can provide valuable information about key processes of planet formation: the planetary accretion, dust growth, and dust population in planet-forming regions. 

Overall, observational surveys looking for planets at wide orbits would minimize circumstellar extinction by targeting low-mass disks over bright disks with substructures with special consideration for inclination effects. This approach would improve the chances of successful campaigns and further characterization of planetary growth.

\begin{acknowledgments}
F.A and E.A.B. acknowledge support from NSF AAG Grant \#1907653. GC thanks the Swiss National Sci-
ence Foundation for financial support under grant number P500PT\_206785.
\end{acknowledgments}

%


\software{\texttt{FARGO3D} \citep{FARGO3D},
\texttt{RADMC-3D} \citep{RADMC-3D},
numpy \citep{numpy},
matplotlib \citep{matplotlib},
cmasher \citep{cmasher}
          }



\appendix \label{Appendix}

\section{Specific Applications}\label{Appendix:disks}

We run multiple specific models for planet-hosting disk candidates targeted in high-contrast imaging campaigns.  To minimize computational cost we used 2D simulations with a  (1024,256) grid in radius and azimuth, which are then expanded vertically assuming hydrostatic equilibrium given the aspect ratio and flaring index values previously reported in the literature. All the values used for each model and their respective reference are listed in Table \ref{Tab:Models}.

For the PA location of the planet, the literature proposes different values depending on the reference. Wherever we chose a specific value we specify them in Table \ref{Tab:Models}. If a specific PA for the planet is not clearly described in the literature we located the planet in the minor axis of the disk by default.

In cases where no constraints of the disk's small dust content are available,  we assume a 10\% of the dust mass for the small grains population and a maximum grain size of $a_{\rm max}$ = 10 $\mu$m. Each dust population has a different scale height for each disk, assuming there is dust settling. The role of dust settling can be significant in some of the most inclined disk cases. The assumed scale height for each dust population is \citep{Johansen..2014}:

\begin{equation}
    H_{\mathrm{dust}} = H_{\mathrm{gas}}\sqrt{\frac{\alpha_D}{\alpha_D + \mathrm{St}}},
\end{equation}

\noindent with St the Stokes number, $\alpha_D$ a diffusion parameter, and $H_{\mathrm{dust}}$ and $H_{\mathrm{gas}}$ the dust and gas scale height respectively. For simplicity, in this approach, we assume that the diffusion viscosity term, $\alpha_D$ is equal to the  $\alpha$ viscosity term listed in Table \ref{Tab:Models} for each disk.  The assumed Stokes numbers for the calculation of the settling are equal to a dust size of $a_{\mathrm{small}}$=10 $\mu$m and $a_{\mathrm{large}}$=1 mm at the radius of the planetary orbit.

Even though the difference in scale height may be relevant for disks with high inclination, the effect of settling should be subtle for the sources with low inclinations as the overlap with the line of sight is not that significant.
We further list the extinction values for the key continuum and hydrogen lines to the most massive planet for each disk in Table \ref{Tab:A_values}.

\begin{center}
 \begin{deluxetable*}{l|ccccccc}
 \caption{Parameter/disk}

\tablehead{
\colhead{} & \colhead{HD 163296} & \colhead{AS 209} & \colhead{PDS 70} & \colhead{TW Hya}  & \colhead{IM Lup} & \colhead{LKCa 15} & \colhead{HD 169142} 
}
\startdata
M$_{*}$ (M$_{\odot}$) & 2.0  & 1.2 & 0.8 & 0.81  & 1.1  & 1.32 & 1.85 \\
M$_{\rm disk}$ (M$_{\odot}$) &  0.14 & 0.0045 &  0.003 & 0.05 & 0.2  & 0.1 & 0.257\\
M$_{\mathrm{mm}}$ (M$_{\odot}$) & 2.31$\times10^{-3}$  & 4.5$\times10^{-4}$ & 3$\times10^{-5}$  & 4$\times10^{-4}$   & 1.97$\times10^{-3}$  &  5$\times10^{-4}$ &  2.6$\times10^{-3}$\\
M$_{\mu \mathrm{m}}$(M$_{\odot}$) & 2.00$\times10^{-4}$   & 5.23$\times10^{-5}$  & 9.7$\times10^{-7}$  & 10$^{-4}$   &2.02$\times10^{-5}$   & 5$\times10^{-5}$ & 2.6$\times10^{-4}$ \\
Aspect ratio($h/r$) & 0.089  & 0.06 & 0.085 & 0.072   & 0.1 & 0.05 & 0.074 \\
Flaring index, $\psi$ & 1.25  & 1.25 & 1.14 & 1.2  & 1.17 & 1.25  & 1.15\\
$\alpha$-viscosity & 10$^{-3}$  & $10^{-4}$  &  $10^{-3}$  & 10$^{-4}$ & 0.03  & $10^{-4}$ & $10^{-3}$ \\
Inclination(\degg) & 46.7  & 35.0 & 49.6 & 7    & 47.5 & 50.2 & 13\\
PA(\degg) & 133  & 85.8 & 158.6 &  155 & 144.5 & 61.9 & 5 \\
Distance (pc) &  101 & 121 & 112 & 60   & 158  & 159  & 114\\
R$_p$ (au) & 48  & 100 & 20.1 & 82  & 127 & 42 & 37 \\
M$_p$ (M$_{\rm Jup}$) & 2.18  & 0.32 & 5 & 0.5  & 0.041 & 0.15 & 1.0 \\
P.A.$_p$ ($^{\circ}$) & -  & - & 155 & 60  & 69 & 60 & 33.9 \\
R$_{p2}$ (au) & 86  & 200 & 33.2 & -  & -  &  -  &  64\\
M$_{p2}$ (M$_{\rm Jup}$) & 0.14  & 0.41 & 2.8 & -   & -  & - &  0.03\\
Refs & 1,2 & 3,4  & 5-9 & 10-12 & 1,13-15  & 16-18  &   19-22
\enddata
\tablerefs{1.\cite{MAPS..V..Coco} 2.\cite{Teague..2022}, 3.\cite{Teague..2018}, 4.\cite{MAPS..VIII..Alarcon},
5.\cite{Keppler..2018},
6.\cite{Keppler..Teague..2019},
7.\cite{Haffert..et..al..2019},
8.\cite{Asensio-Torres..21},
9.\cite{Portilla_Revelo..2022},
10.\cite{Huang..2018a},
11.\cite{Calahan..2021},
12.\cite{Teague..2022},
13.\cite{Speedie..22},
14.\cite{Rosotti..23},
15.\cite{Pinte..2020},
16.\cite{Long..22},
17.\cite{Facchini..2020},
18.\cite{Jin..2019},
19.\cite{Perez..2020},
20.\cite{Wang..21},
21.\cite{Garg..22},
22.\cite{Hammond..2023}}
\label{Tab:Models}
\end{deluxetable*}
\end{center}

\begin{center}
 \begin{deluxetable*}{l|ccccccc}
 \caption{Extinction values for putative protoplanets}
\label{Tab:A_values}
\tablehead{
\colhead{Band} & \colhead{HD 163296} & \colhead{AS 209} & \colhead{PDS 70} & \colhead{TW Hya}  & \colhead{IM Lup} & \colhead{LKCa 15} & \colhead{HD 169142} 
}
\startdata
H & 62.7  & 8.6 & 2.8 & 29.8  & 16.4  & 13.8 & 28.1 \\
K & 52.5  & 7.1 & 2.3 & 24.8  & 13.7  & 11.5 & 23.5 \\
L & 40.7  & 5.1 & 1.8 & 18.6  & 11.0  & 8.9 & 18.0 \\
M & 34.9  & 4.3 & 1.6 & 15.7  & 9.4  & 7.6 & 15.4 \\
N1 & 25.7  & 3.0 & 1.2 & 11.4  & 7.0  & 5.6 & 11.3 \\
N2 & 29.5  & 3.8 & 1.3 & 13.7  & 7.8  & 6.5 & 13.1 \\
\hline 
H$\alpha$ & 100.6  & 14.2 & 4.4 & 48.6  & 26.0  & 22.2 & 45.3 \\
Pa$\alpha$ & 57.4  & 7.8 & 2.5 & 27.2  & 15.0  & 12.6 & 25.7 \\
Br$\alpha$ & 38.9  & 4.9 & 1.7 & 17.7  & 10.4  & 8.5 & 17.3 \\
Br$\beta$ & 47.2  & 6.4 & 2.1 & 22.4 & 12.4  & 10.4 & 21.2 \\
\enddata
\end{deluxetable*}
\end{center}

\begin{figure}
    \centering
    \includegraphics[width=0.99\linewidth]{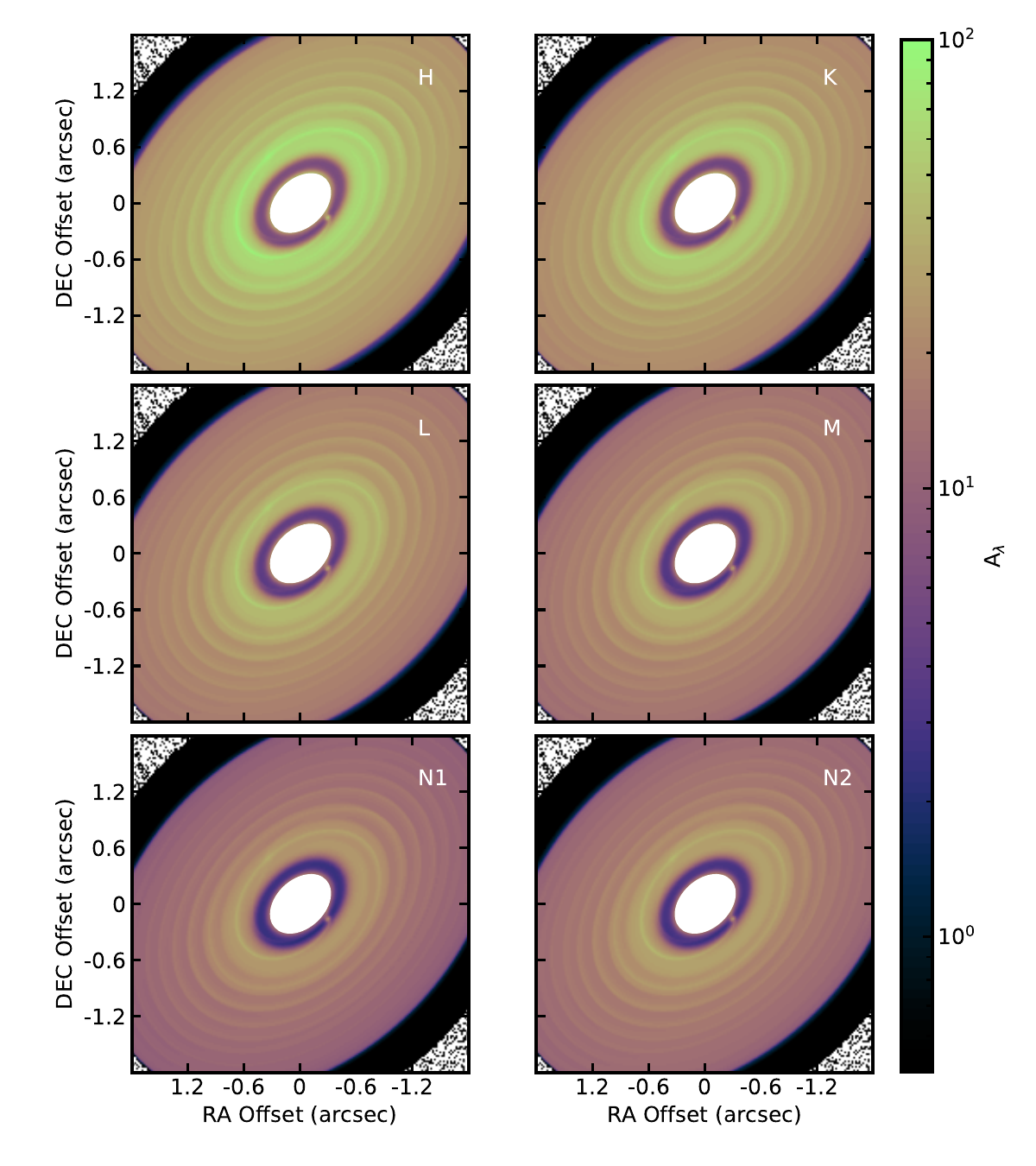}
    \includegraphics[width=0.99\linewidth]{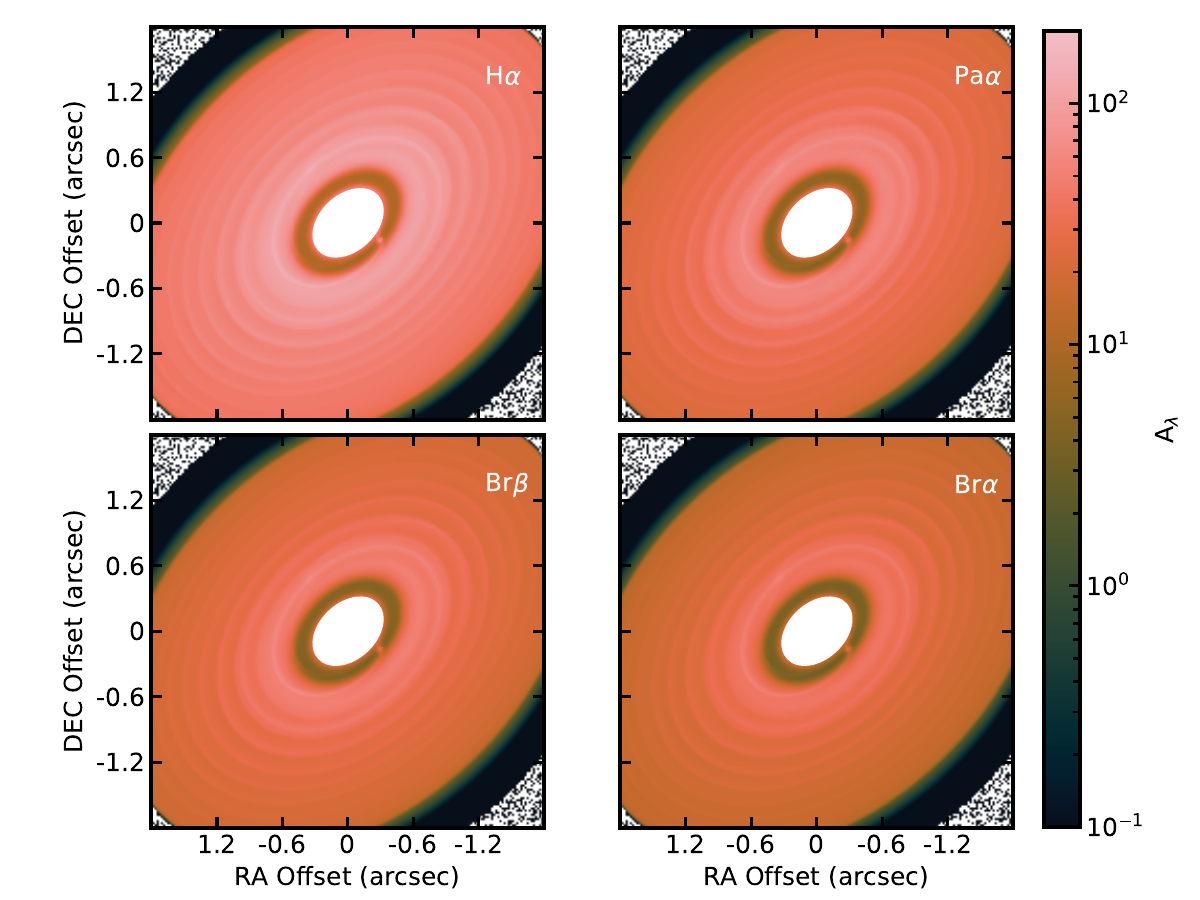}
    \caption{Extinction from HD 163296 model.}
    \label{fig:enter-label}
\end{figure}

\begin{figure}
    \centering
    \includegraphics[width=0.99\linewidth]{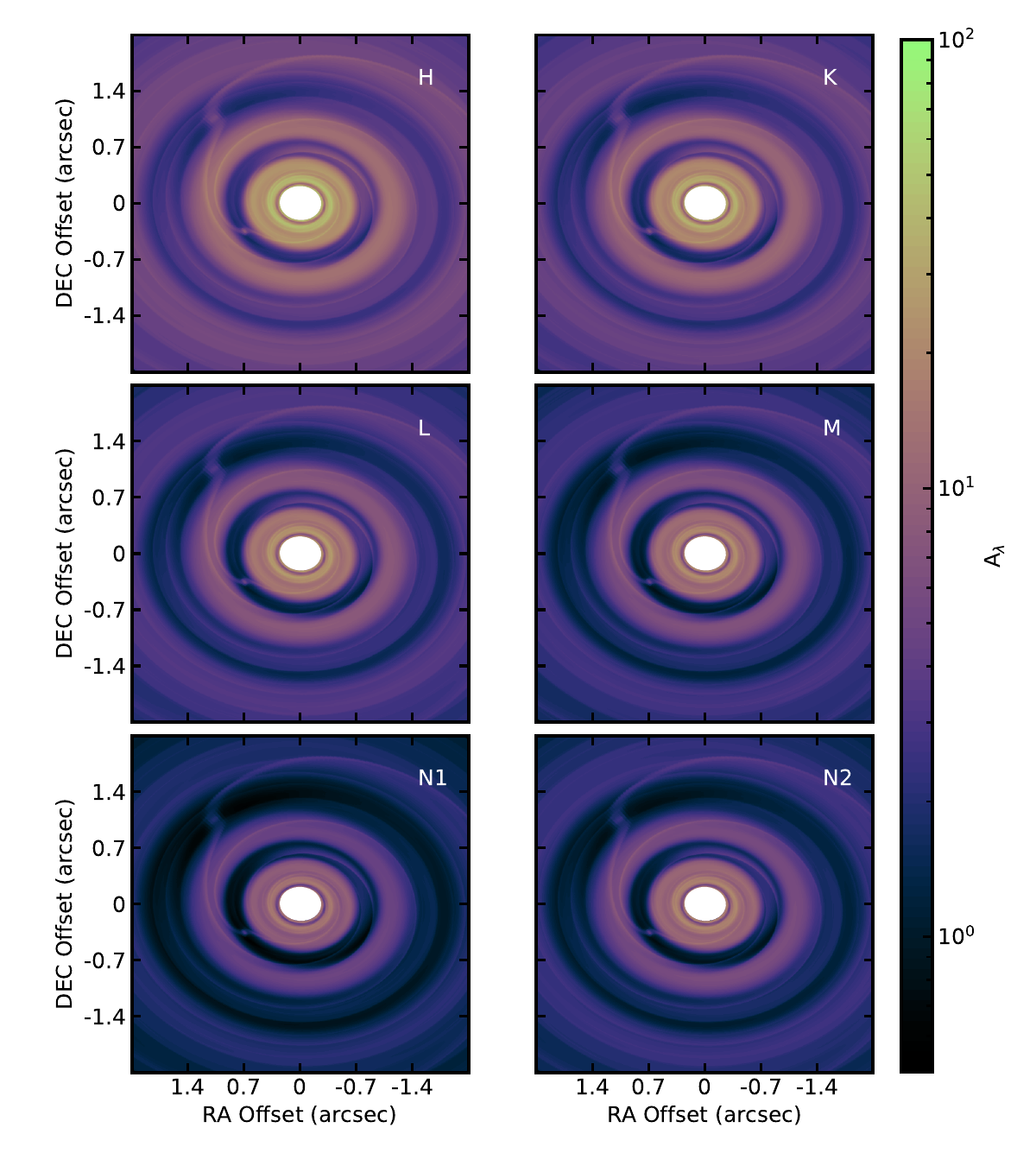}
    \includegraphics[width=0.99\linewidth]{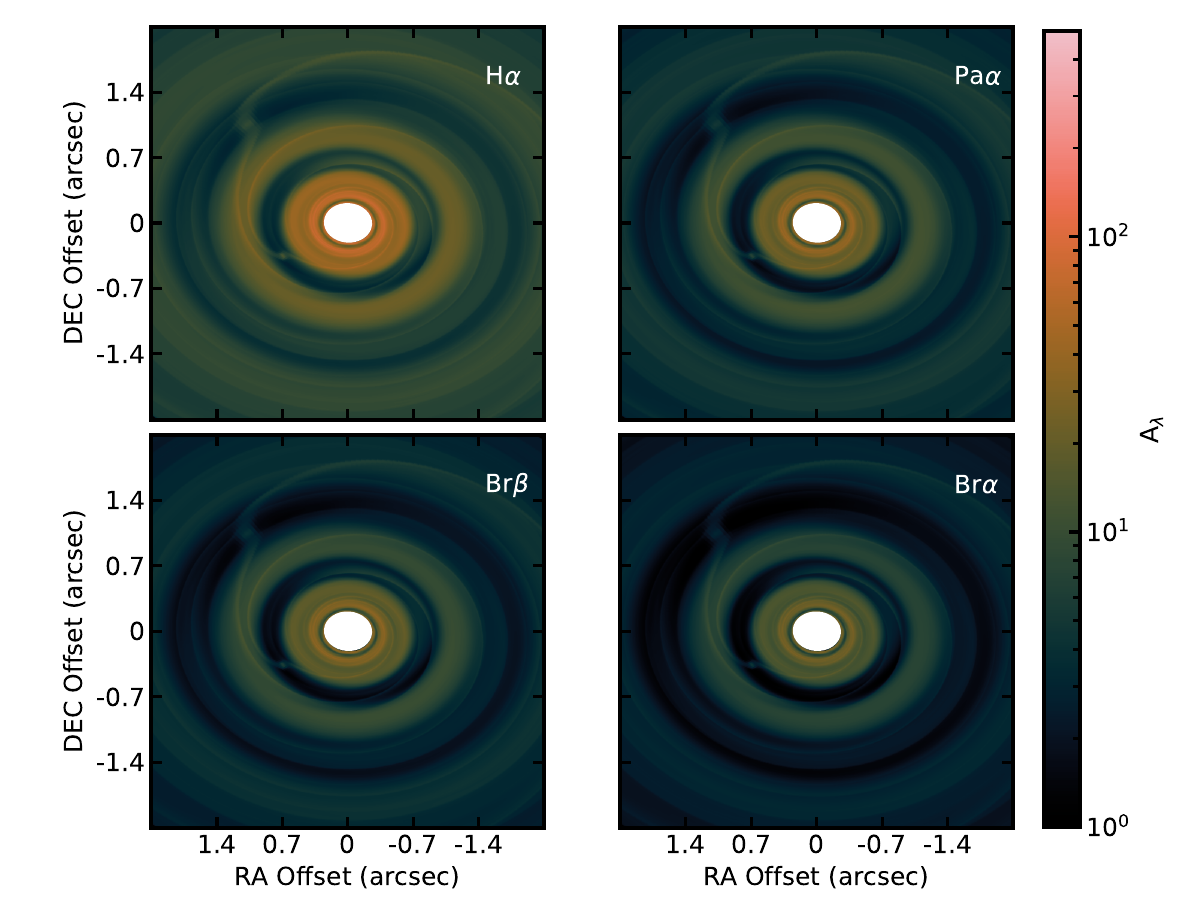}
    \caption{Extinction from AS 209 model.}
    \label{fig:enter-label}
\end{figure}


\begin{figure}
    \centering
    \includegraphics[width=0.99\linewidth]{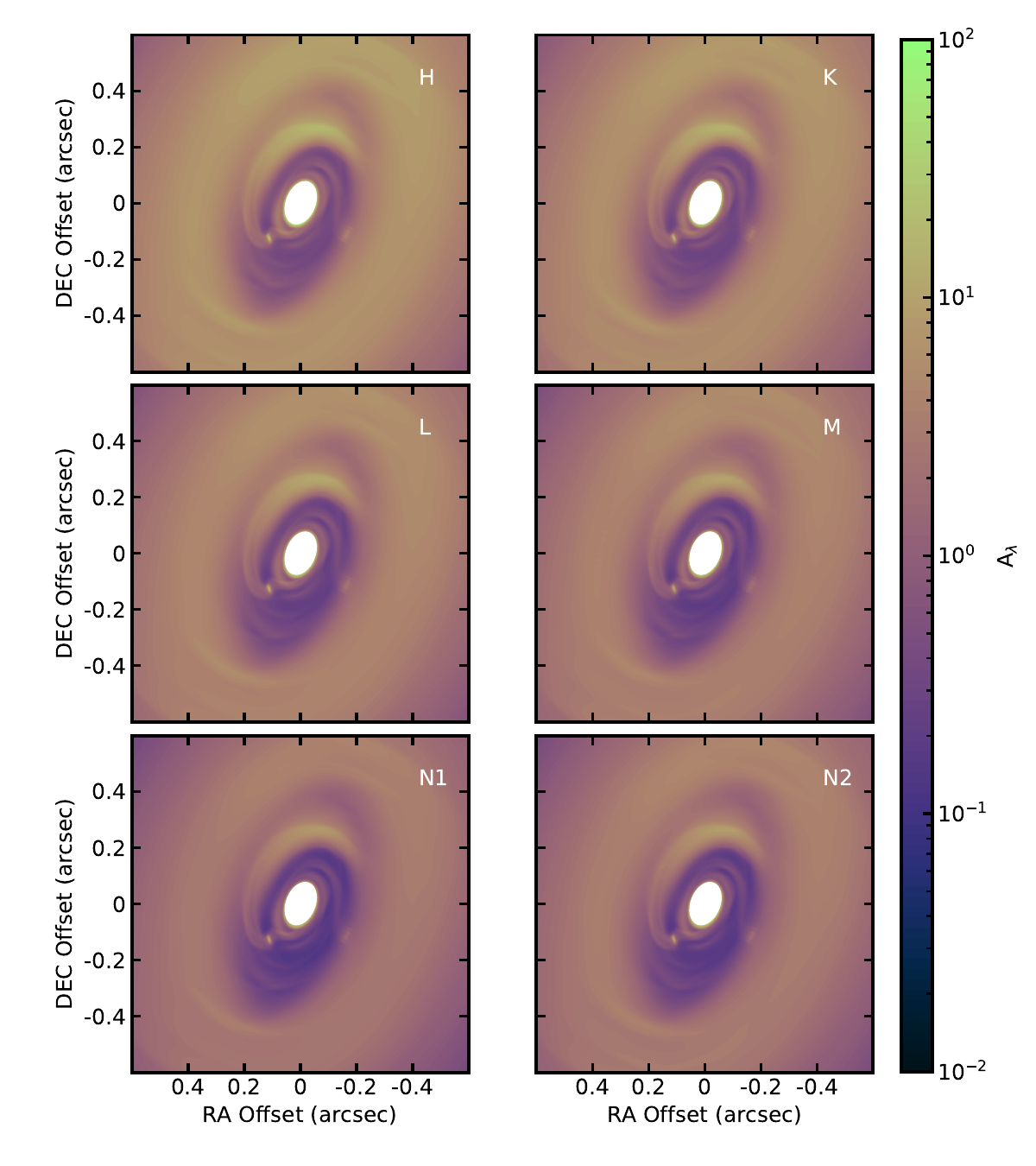}
    \includegraphics[width=0.99\linewidth]{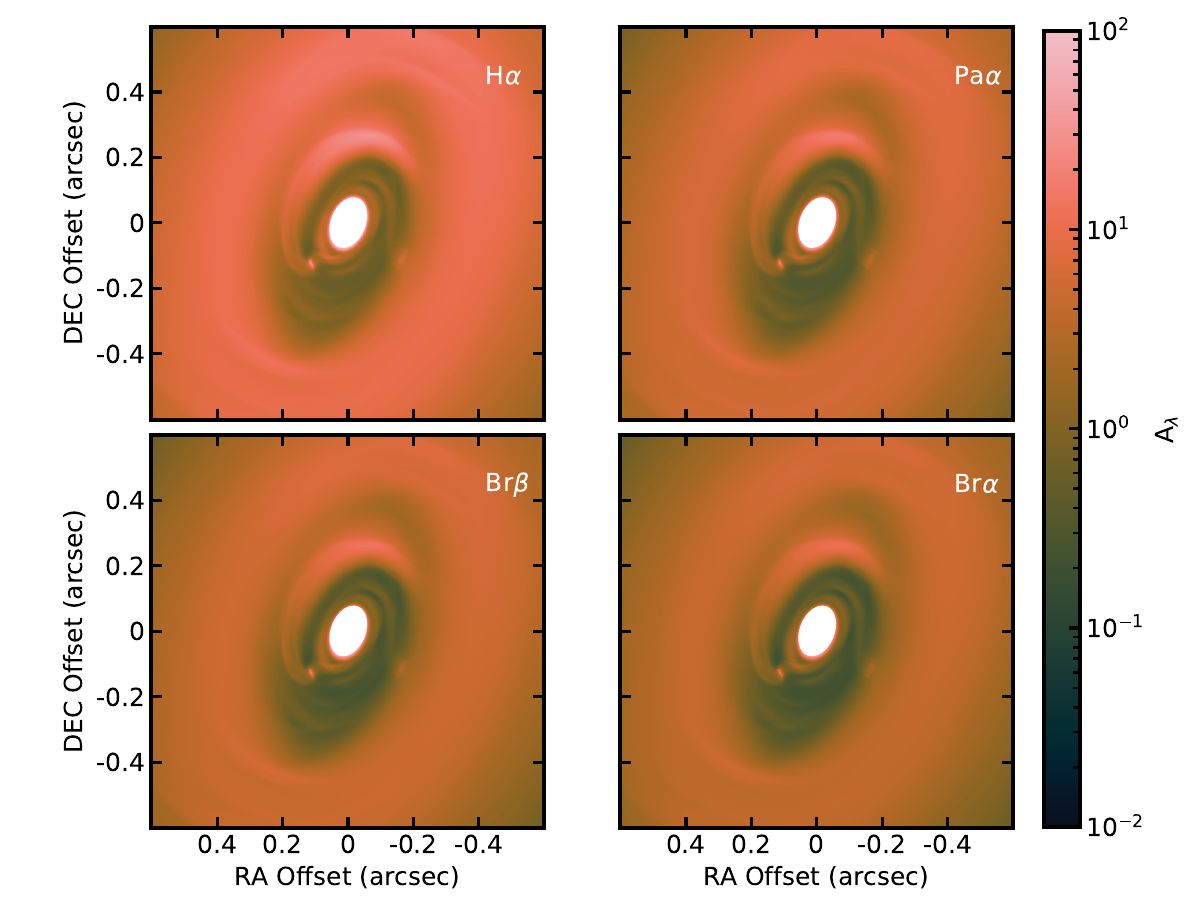}
    \caption{Extinction from PDS 70 model.}
    \label{fig:enter-label}
\end{figure}

\begin{figure}
    \centering
    \includegraphics[width=0.99\linewidth]{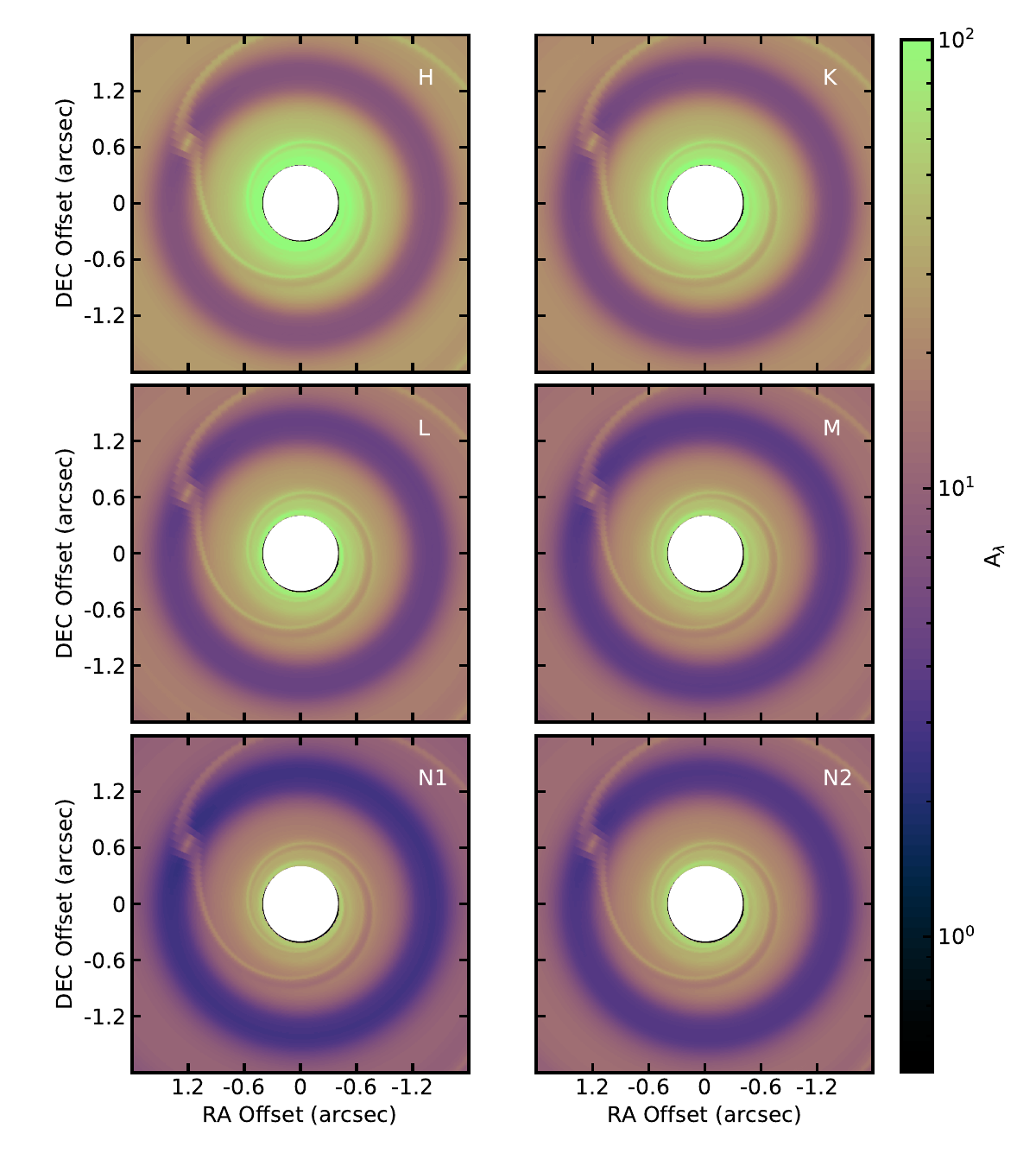}
    \includegraphics[width=0.99\linewidth]{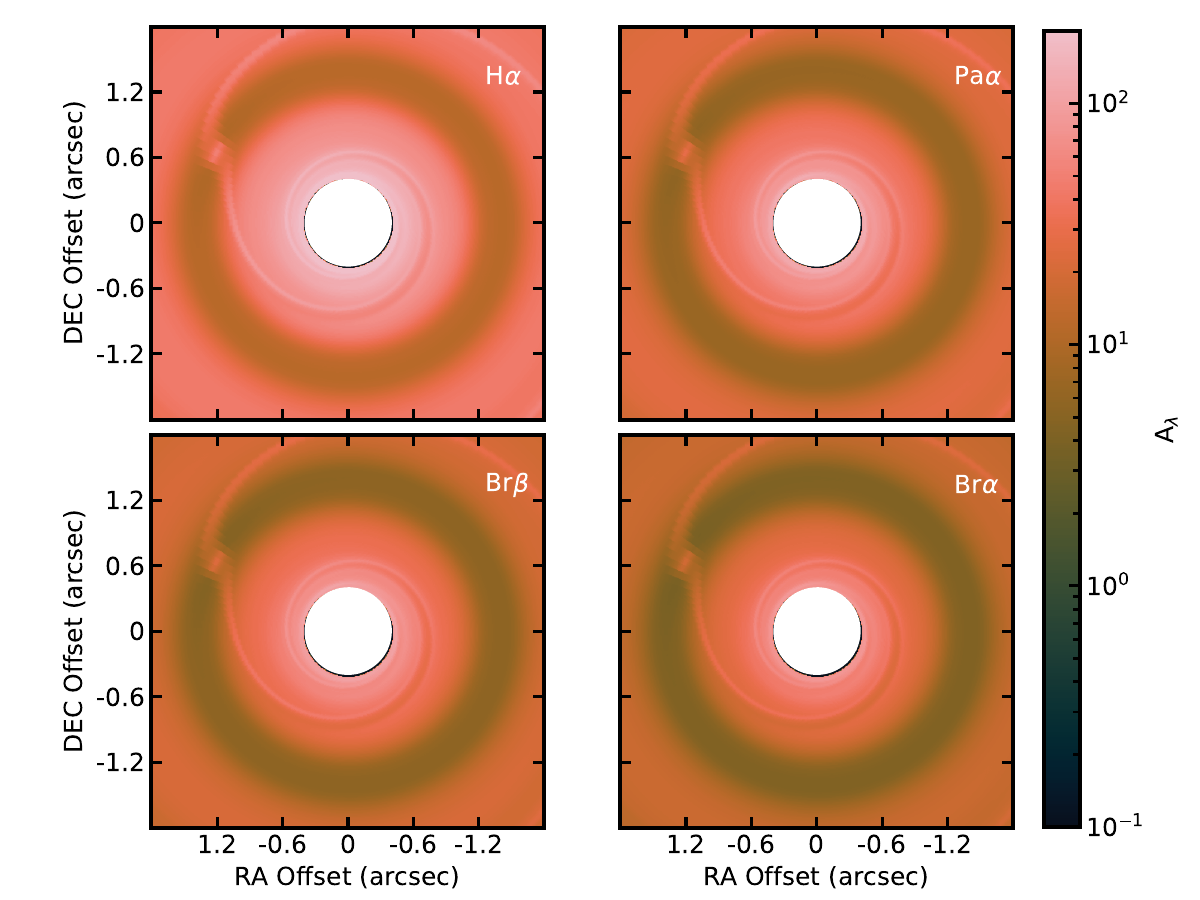}
    \caption{Extinction from TW Hya model.}
    \label{fig:enter-label}
\end{figure}

\begin{figure}
    \centering
    \includegraphics[width=0.99\linewidth]{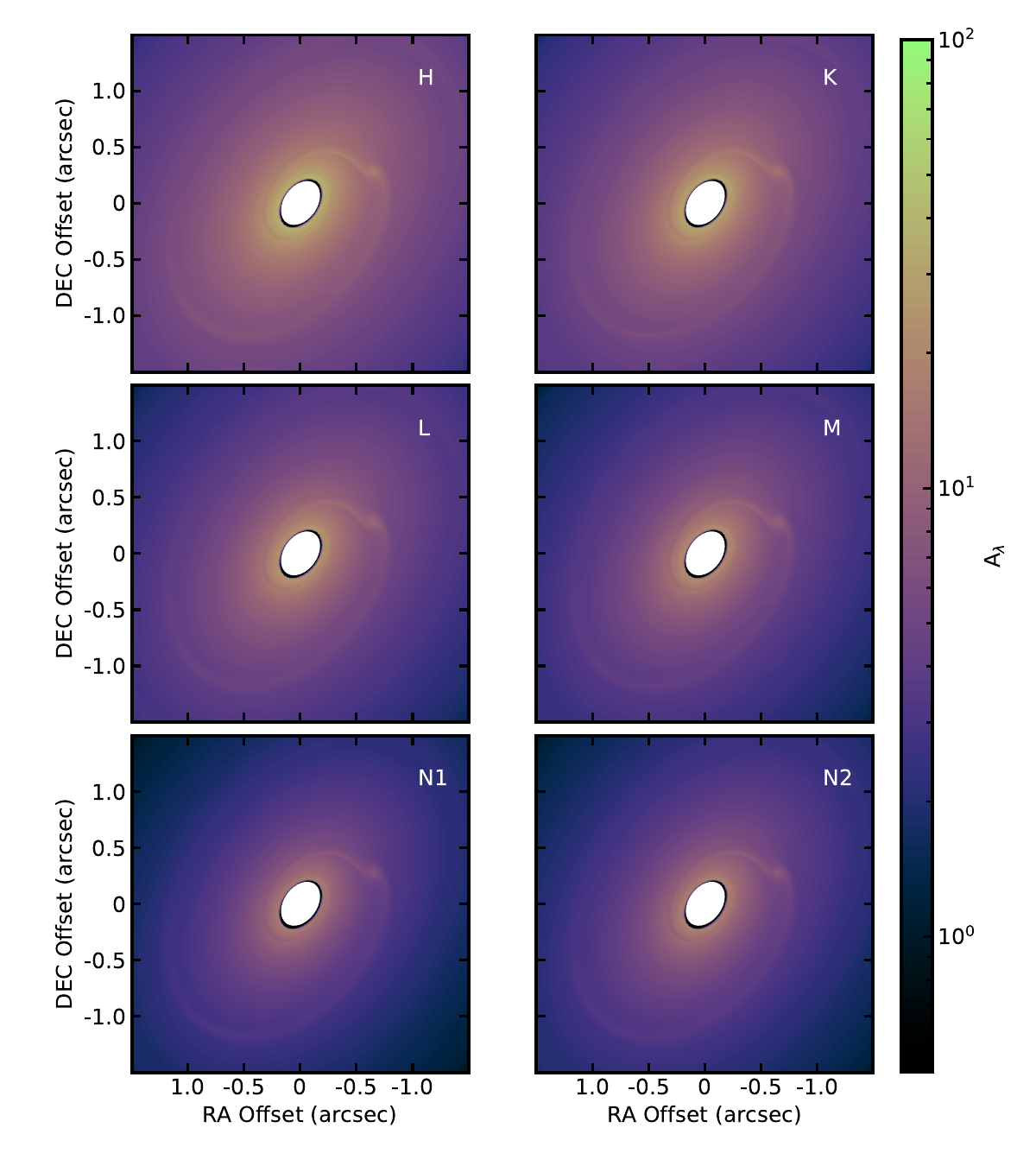}
    \includegraphics[width=0.99\linewidth]{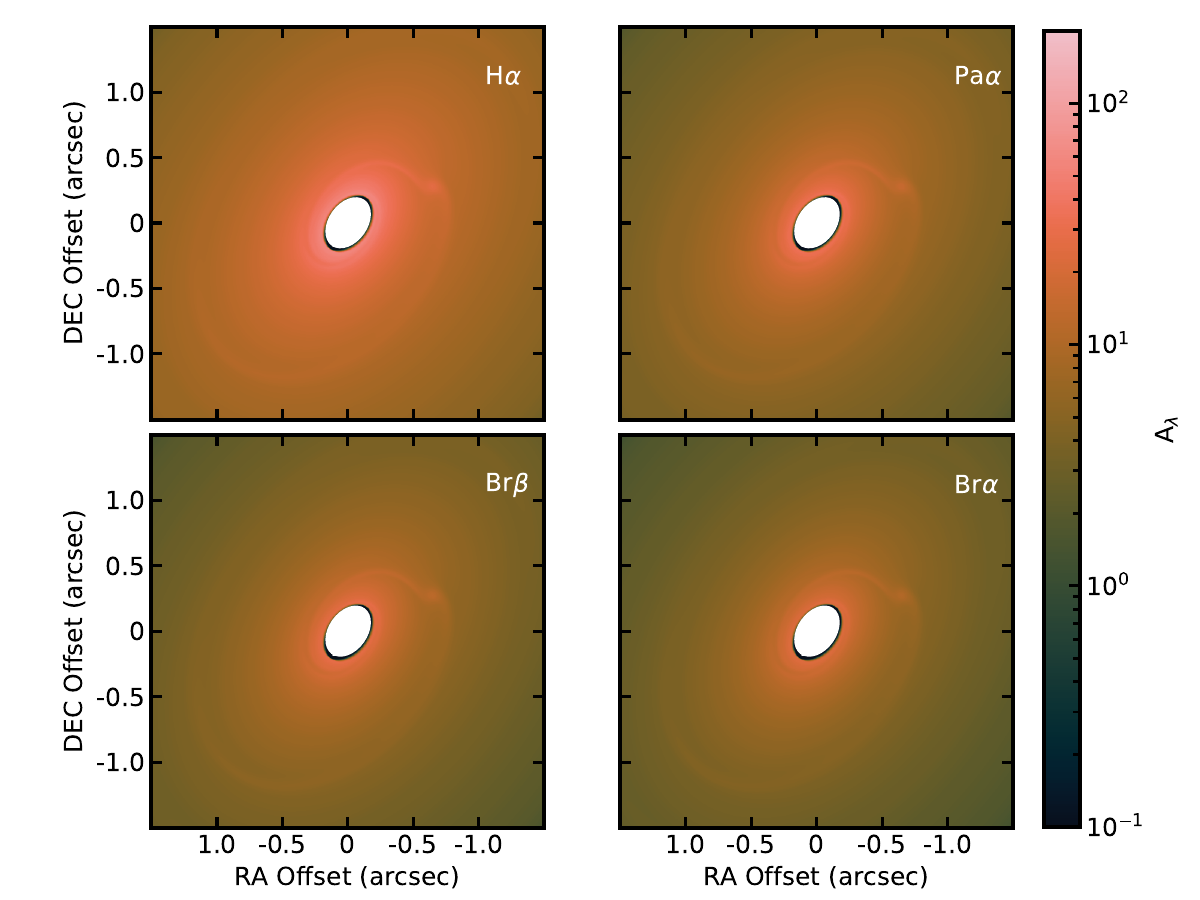}
    \caption{Extinction from IM Lup model.}
    \label{fig:enter-label}
\end{figure}

\begin{figure}
    \centering
    \includegraphics[width=0.99\linewidth]{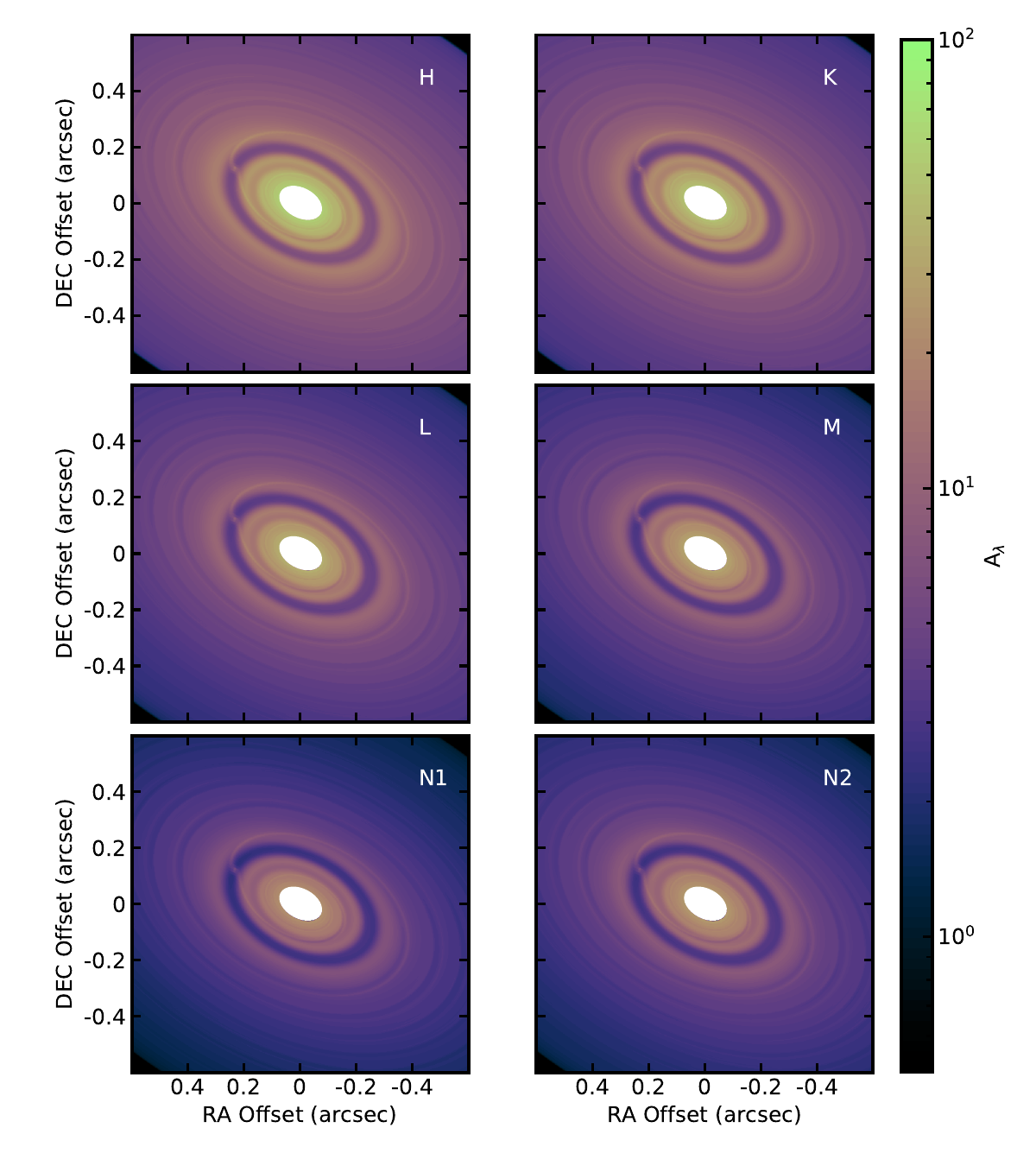}
    \includegraphics[width=0.99\linewidth]{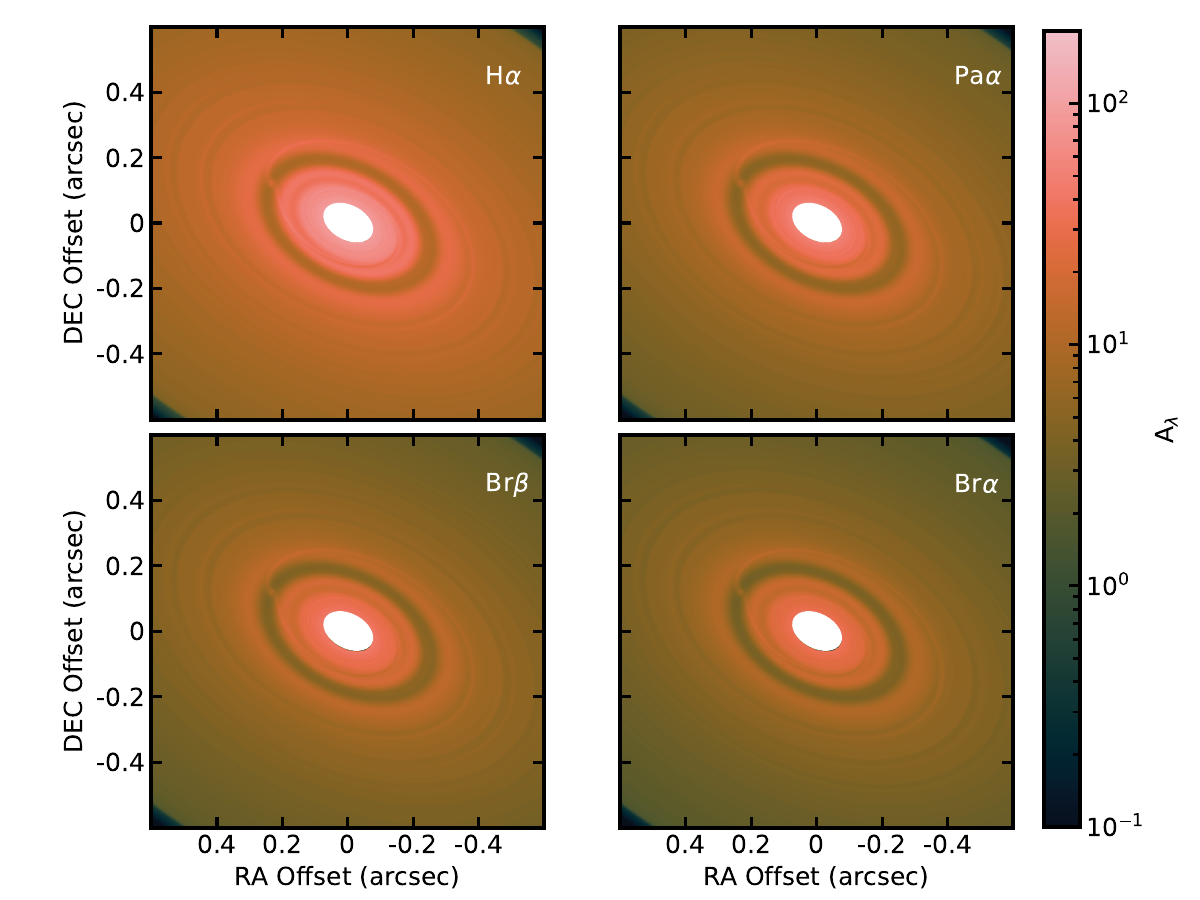}
    \caption{Extinction from LkCa 15 model.}
    \label{fig:enter-label}
\end{figure}

\begin{figure}
    \centering
    \includegraphics[width=0.99\linewidth]{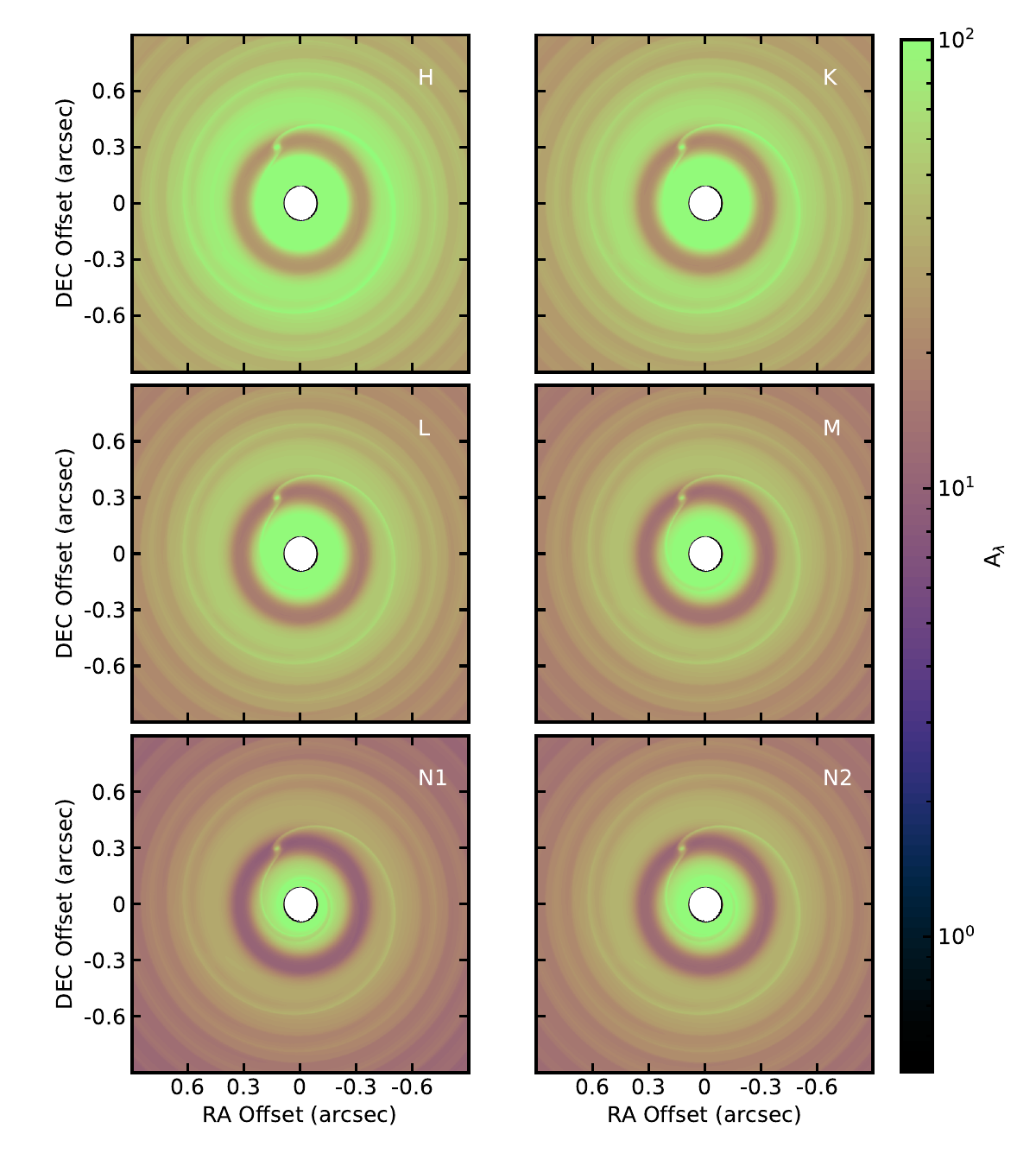}
    \includegraphics[width=0.99\linewidth]{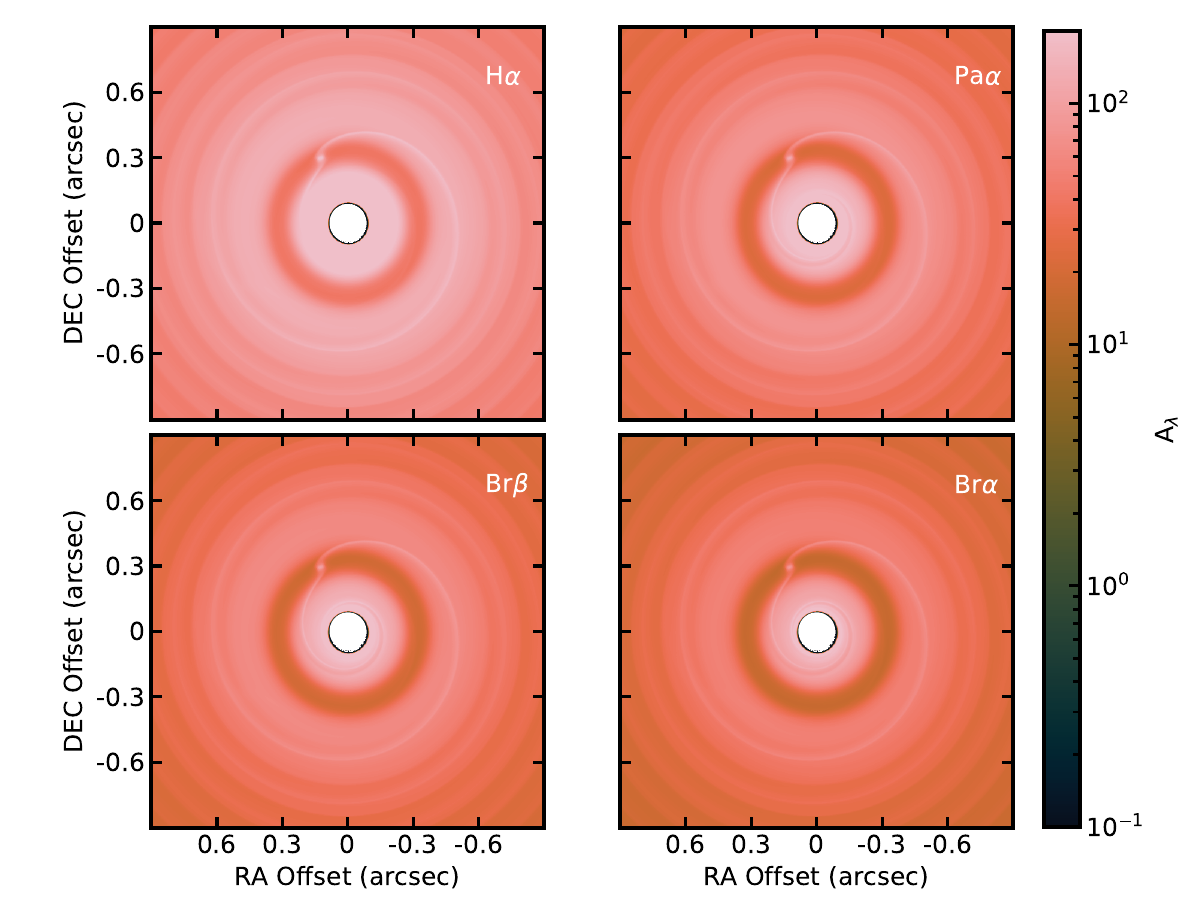}
    \caption{Extinction from HD 169142 model.}
    \label{fig:enter-label}
\end{figure}

\section{Variation of the extinction values along the protoplanet's orbit}\label{Appendix:PA_devs}

\begin{figure*}
    \centering
    \includegraphics[width=0.99\linewidth]{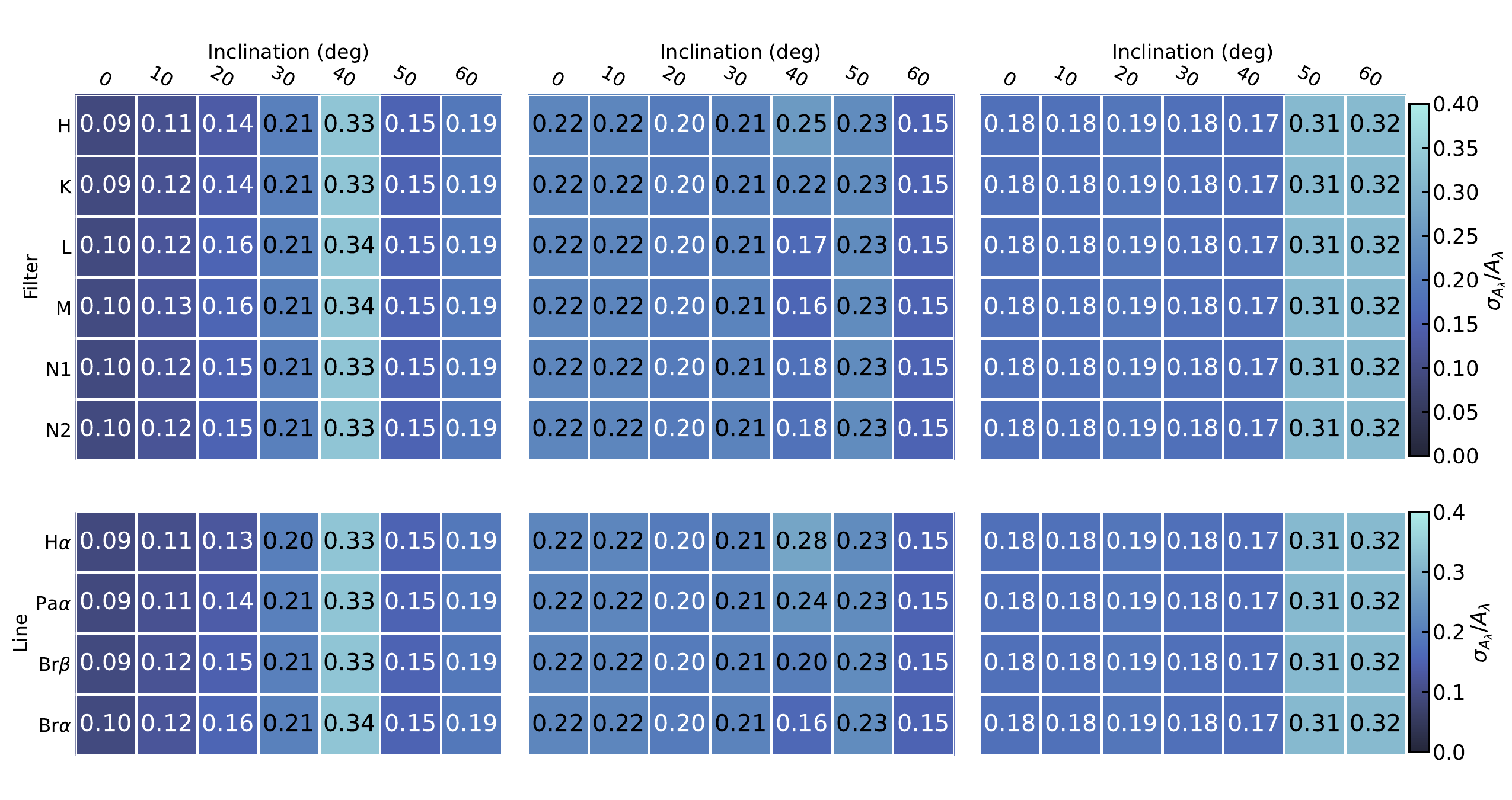}
    \caption{Relative standard deviations for the extinction in the protoplanetary orbit for the different geometries and inclinations.}
    \label{fig:PA_vars}
\end{figure*}

For inclined disks, the amount of extinction towards the protoplanet will depend on the position angles it has in its orbit. Depending on the azimuth it is located, it may have a larger angular separation or more circumstellar material along the line of sight. In all our models we assumed that the protoplanet is in the minor axis of the disk. We explore the possible variation the extinction may go through depending on the protoplanet locations.  We show the ratio of the standard deviation variation across the protoplanet's orbit with respect to the nominal values in Figure \ref{fig:PA_vars}. For more inclined disks, those variations are the largest reaching levels up to 30\%. The relative ratio increases with planetary masses because the nominal values for the Jovian and super-Jovian planets produce low extinction, so in absolute values, the range is actually smaller. For face-on and low-inclination disks, the variations are mostly induced by the spiral wakes of the planet.

\bibliography{ref}{}
\bibliographystyle{aasjournal}



\end{document}